%% file: 2PZS-LQ-MFTG.tex
\documentclass[final]{l4dc2024}
\usepackage[utf8]{inputenc}

\usepackage{amssymb,amsmath}
\usepackage{soul}
\usepackage{graphicx}
\usepackage{caption}
\usepackage{subcaption}
\usepackage[left=2cm,right=2cm,top=2cm,bottom=2cm]{geometry}
\usepackage{xcolor}
\usepackage{bbm}
\usepackage{algorithm}
\usepackage{algorithmic}
\usepackage{epstopdf}
\usepackage{array}
\usepackage{times}
\usepackage{comment}
\usepackage{boondox-cal}
\usepackage{subcaption}

\newtheorem*{proposition*}{Proposition}
\newtheorem*{theorem*}{Theorem}

\newenvironment{proof}
{\textit{Proof:} }
{$\square$}

\newcommand{\EE}{\mathbb{E}}

\newcommand{\Ns}{\mathcal{N}}

\newcommand{\Us}{\mathcal{U}}
\newcommand{\Os}{\mathcal{O}}

\newcommand{\RR}{\mathbb{R}}

\newcommand{\Sb}{\mathbb{S}}

\newcommand{\bA}{\bar{A}}

\newcommand{\bc}{\bar{c}}

\newcommand{\bx}{\bar{x}}
\newcommand{\bu}{\bar{u}}
\newcommand{\bB}{\bar{B}}
\newcommand{\bQ}{\bar{Q}}
\newcommand{\bL}{\bar{L}}
\newcommand{\bM}{\bar{M}}
\newcommand{\bR}{\bar{R}}
\newcommand{\bK}{\bar{K}}
\newcommand{\bP}{\bar{P}}
\newcommand{\bN}{\bar{N}}
\newcommand{\bomega}{\bar{\omega}}
\newcommand{\bSigma}{\bar{\Sigma}}
\newcommand{\bLambda}{\bar{\Lambda}}
\newcommand{\bPsi}{\bar{\Psi}}
\newcommand{\mcu}{\mathcal{u}}

\newcommand{\tA}{\tilde{A}}

\newcommand{\tB}{\tilde{B}}

\newcommand{\tP}{\tilde{P}}
\newcommand{\tM}{\tilde{M}}
\newcommand{\tK}{\tilde{K}}
\newcommand{\tL}{\tilde{L}}
\newcommand{\tbK}{\tilde{\bar{K}}}
\newcommand{\tbL}{\tilde{\bar{L}}}
\newcommand{\tz}{\tilde{z}}
\newcommand{\tJ}{\tilde{J}}

\newcommand{\tomega}{\tilde{\omega}}
\newcommand{\tSigma}{\tilde{\Sigma}}
\newcommand{\tnabla}{\tilde{\nabla}}

\newcommand{\ugamma}{\underline{\gamma}}

\newcommand{\hK}{\hat{K}}
\newcommand{\hLambda}{\hat{\Lambda}}

\DeclareMathOperator{\tr}{Tr}

\DeclareMathOperator*{\argmin}{argmin}

\DeclareMathOperator{\diag}{diag} 

\DeclareMathOperator{\proj}{Proj}

\definecolor{NavyBlue}{rgb}{0.0, 0.0, 0.5}

\newcommand{\verti}[1]{{\left\vert\kern-0.25ex\left\vert\kern-0.25ex\left\vert #1 
    \right\vert\kern-0.25ex\right\vert\kern-0.25ex\right\vert}}

\title[Robust Cooperative MARL]{Robust Cooperative Multi-Agent Reinforcement Learning:\\ A Mean-Field Type Game Perspective}
\author{%
 \Name{Muhammad Aneeq uz Zaman} \Email{mazaman2@illinois.edu}\\
 \addr Coordinated Science Laboratory, University of Illinois at Urbana-Champaign
 \AND
 \Name{Mathieu Laurière} \Email{mathieu.lauriere@nyu.edu}\\
 \addr Shanghai Frontiers Science Center of Artificial Intelligence and Deep Learning;\\
 NYU-ECNU Institute of Mathematical Sciences at NYU Shanghai 
 \AND
 \Name{Alec Koppel} \Email{alec.koppel@jpmchase.com}\\
 \addr Artificial Intelligence Research, JP Morgan Chase \& Co.
 \AND
 \Name{Tamer Ba{\c s}ar} \Email{basar1@illinois.edu}\\
 \addr Coordinated Science Laboratory, University of Illinois at Urbana-Champaign
}

\begin{document}
	
	\maketitle
	\begin{abstract}
        In this paper, we study the problem of robust cooperative multi-agent reinforcement learning (RL) where a large number of cooperative agents with distributed information aim to learn policies in the presence of \emph{stochastic} and \emph{non-stochastic} uncertainties whose distributions are respectively known and unknown. Focusing on policy optimization that accounts for both types of uncertainties, we formulate the problem in a worst-case (minimax) framework, which is is intractable in general. Thus, we focus on the Linear Quadratic setting to derive  benchmark solutions. First, since no standard theory exists for this problem due to the distributed information structure, we utilize the Mean-Field Type Game (MFTG) paradigm to establish guarantees on the solution quality in the sense of achieved Nash equilibrium of the MFTG. This in turn allows us to compare the performance against the corresponding original robust multi-agent control problem. Then, we propose a Receding-horizon Gradient Descent Ascent RL algorithm to find the MFTG Nash equilibrium and we prove a non-asymptotic rate of convergence. Finally, we provide numerical experiments to demonstrate the efficacy of our approach relative to a baseline algorithm.
	\end{abstract}
\section{Introduction}
Reinforcement Learning (RL) has had many successes, such as autonomous driving \citep{sallab2017deep},  
robotics \citep{kober2013reinforcement}, and RL from human feedback (RLHF) \citep{ziegler2019fine}, to name a few. These successes have been focused on single-agent scenarios, but many scenarios involving,  e.g.,  financial markets, communication networks, distributed robotics involve multiple agents. Prevailing algorithms for Multi-Agent Reinforcement Learning (MARL) \citep{zhang2021multi,li2021distributed}, however, do not model the distinct effects of modeled and un-modeled uncertainties on the transition dynamics, which can result in practical instability in safety-critical applications \citep{riley2021utilising}.

In this paper we consider a large population multi-agent setting, with stochastic and non-stochastic (un-modeled, possibly adversarial) uncertainties. These types of formulations have been studied under the guise of robust control in the single-agent case \citep{bacsar2008h}. The uncertainties (modeled and un-modeled) affect the performance of the system and might even lead to instability. Robust control seeks the \emph{robust} controller which guarantees a certain level of performance for the system in under a worst-case hypothesis on these uncertainties. 
We employ here the popular Linear-Quadratic (LQ) setting in order to rigorously characterize and synthesize the solution to the robust multi-agent problem in a data-driven manner. The LQ setting entails a class of models in which the dynamics are linear and the costs are quadratic in the state and the action of the agent. This setting has been used extensively in the literature due to its tractability: the optimal decisions can be computed analytically or almost analytically, up to solving Riccati equations, when one has access to all system matrices. Instances of applications include permanent income theory \citep{sargent2000recursive}, portfolio management \citep{cardaliaguet2018mean}, and wireless power control \citep{huang2003individual}), among many others. In the absence of knowledge of system parameters, model-free RL methods have also been developed \citep{fazel2018global,malik2019derivative} for single agent LQ settings. We refer to~\citep{recht2019tour} for an overview. 
When one goes from single to multiple agents, the issue of communicating local state and control information among agents exhibit scalability problems, and in particular, practical algorithms require sharing state information that can scale exponential in the number of agents. Instead, here we consider a distributed information structure where each agent has access only to its own state and the average of states of the other agents. This distributed information structure causes the characterization of the solution to be very difficult, in that previous gradient dominance results from \citep{fazel2018global} no longer hold. 
To overcome this difficulty, we utilize the mean-field game and control paradigm, first introduced in the purely non-cooperative agent setting in \citep{lasry2006jeux,huang2006large}, which replaces individual agents by a distribution over agent types, which enables characterization and computation of the solution. 
The approach has then been extended to the cooperative setting through the notion of mean field control~\citep{bensoussan2013mean,carmona2018prob}. Building on this paradigm, this work is the first to develop scalable algorithms for MARL   that can handle model mis-specification or adversarial inputs in the sense of robust control in the very large or possibly infinite number of agents defined by the mean-field.  





We start Section \ref{sec:form} by formulating a robust multi-agent control problem with stochastic and non-stochastic (un-modeled) noises. The agents have distributed information, such that they have access to their own states and the average behavior of all the agents. Solving this problem entails finding a \emph{noise attenuation level} (noise-to-output gain) for the multi-agent system and the corresponding \emph{robust controller}. As in the single-agent setting \citep{bacsar2008h}, the robust multi-agent control problem is reformulated into an equivalent zero-sum min-max game between the maximizing non-stochastic noise (which may be interpreted as an adversary) and the minimizing controller. Solving this problem is not possible in the finite agent case due to the limited information available to each agent. Thus, in Section \ref{sec:RMFC} we consider the mean-field (infinite population) version of the problem, that we call the \emph{Robust Mean-Field Control} (RMFC) problem. As in the finite-population setting, RMFC has an equivalent zero-sum min-max formulation, referred to as the 2-player \emph{Zero-Sum Mean-Field Type Game} (ZS-MFTG) in 
\citep{carmona2020policy,carmona2021linear}, where the controller is the minimizing player and the non-stochastic disturbance is the maximizing one. 

In Section \ref{sec:RL_MFTG} we propose a bi-level RL algorithm to compute the Nash equilibrium for the ZS-MFTG (which equivalently yields the robust controller for the robust multi-agent problem) in the form of \emph{Receding-horizon Gradient Descent Ascent} (RGDA)  (Algorithm \ref{alg:RL^2_tP_MFTG}). 
The upper-level of RGDA, uses a receding-horizon approach, i.e., it finds the controller parameters starting from the last timestep $T-1$ and moving backwards-in-time (à la dynamic programming). The receding-horizon policy gradient approach was used in Kalman filtering \citep{zhang2023learning} and LQR problems \citep{zhang2023revisiting}. The present work builds on this approach to multi-agent problems, which helps in simplifying the complex nature of the cost landscape (known to be non-coercive \citep{zhang2021policy}) and renders it convex-concave. The lower-level employs gradient descent-ascent to find the saddle point (Nash equilibrium) for each timestep $t$. The convex-concave nature of the cost (due to the receding-horizon approach) proves to be a key component in proving linear convergence of the gradient descent-ascent to the saddle point (Theorem~\ref{thm:inner_loop_conv}). 
Further analysis shows that the total accumulated error in the RGDA is small given that the lower level of RGDA has good convergence (Theorem~\ref{thm:main_res}). The gradient descent-ascent step requires computation of the stochastic gradient. We use a zero-order method \citep{fazel2018global,malik2019derivative} which only requires access to the cost to compute stochastic gradients, and hence is \emph{truly} model-free. 

\textbf{Literature Review:} Robust control gained importance in the 1970s when control theorists realized the shortcomings of optimal control theory in dealing with model uncertainties \citep{athans1977stochastic, harvey1978quadratic}. The work of \citep{bacsar1989dynamic} was the first one to formulate the robust control problem as a zero-sum dynamic game between the controller and the uncertainty. Robust RL first introduced by \citep{morimoto2005robust} has recently had an increase in interest in for the single agent setting, where its ability to process trajectory data without explicit knowledge of system parameters can be used to learn robust controllers to address worst-case uncertainty \citep{zhang2020brobust,kos2017delving,zhang2021derivative}. 
Some recent works consider RL in scenarios with reward uncertainties \citep{zhang2020arobust}, state uncertainty \citep{he2023robust} or uncertainty in other agents' policies \citep{sun2022romax}. There have been some works on the intersection of RL for robust and multi-agent control \citep{li2019robust,he2023robust}, yet there has not 
been any significant effort to provide (1) sufficient conditions for solvability of the multi-agent robust control problem i.e. determining the noise attenuation level of a system and (2) provable Robust multi-agent RL (RMARL) algorithms in the large population setting, as proposed in this paper.


This is made possible due to the mean-field game and control paradigm, which considers the limiting case as the number of agents approaches infinity. This paradigm was first introduced in the context of non-cooperative game theory as Mean-Field Games (MFGs) concurrently by \citep{lasry2006jeux,huang2006large}. 
Since then, the question of learning equilibria in MFGs has gained momentum, see~\citep{lauriere2022learning}. In particular, there have been several works dealing with RL for MFGs \citep{guo2019learning,elie2020convergence,perrin2020fictitious,zaman2020reinforcement,xie2021learning,anahtarci2023q}, deep RL for MFGs~\citep{perrin2021mean,cui2021approximately,lauriere2022scalable}, learning in multi-population MFGs \citep{perolat2022scaling,zaman2021adversarial,uz2023reinforcement}, independent learning in MFGs \citep{yongacoglu2022independent,yardim2023policy}, oracle-free RL for MFGs \citep{angiuli2022unified,zaman2023oracle} and RL for graphon games \citep{cui2021learning,fabian2023learning}. There have also been several works on RL for MFC, which is the cooperative counterpart, see e.g.~\citep{carmona2019linear,carmona2019model,gu2021mean,mondal2022approximation,angiuli2022unified}. But these works require ability to sample from the true transition model, and hence are inapplicable in the case of mis-specification or modeling errors. 
To address this setting, we introduce the Robust MFC problem. We will connect this problem to MFTGs~\cite{tembine2017mean}, which contain mixed cooperative-competitive elements. Zero-sum MFTG model a zero-sum competition between two infinitely large teams of agents.
Prior work on the theoretical framework of zero-sum MFTG include~\citep{choutri2019optimal,tembine2017mean,cosso2019zero,carmona2021linear,guan2024zero}. Related to RL, the works \citep{carmona2020policy,carmona2021linear} propose a data-driven RL algorithm based on Policy Gradient to compute the Nash equilibrium between the two coalitions in an LQ setting but do not provide a theoretical analysis of the algorithm. 


\section{Formulation} \label{sec:form}
In this section we introduce the robust multi-agent control problem by first defining the dynamics of the multi-agent system along with its performance and noise indices. The performance and noise indices have been introduced in the literature \citep{bacsar2008h} in order to quantify the affect of the accumulated noise (referred to as noise index) on the performance of the system (called the performance index).  
The noise attenuation level is then defined as an upper bound on the ratio between the performance and noise indices given that the agents employ a robust controller. Hence the robust multi-agent problem is that of finding the robust controller under which a certain noise attenuation is achieved. In order to solve this problem, we reformulate it as a min-max game problem as in the single-agent setting \citep{bacsar2008h}.
Consider an $N$ agent system. We let $[N] = \{1,\dots,N\}$. The $i^{th}$ agent has dynamics which are linear in its state $x^i_t \in \RR^m$, its action $u^{1,i}_t \in \RR^p$, and the mean-field counterparts, $\bar{x}_t$ and $\bar{u}^{i}_t$. 
The disturbance $u^{2,i}_t$ is referred to as \emph{non-stochastic} noise\footnote{The non-stochastic noise is assumed to have identity coefficient in the dynamics \eqref{eq:RMS_dyn} for simplicity of analysis but can be easily changed to some other matrix of appropriate size.} since it is an un-modeled disturbance and can even be adversarial. This is similar in spirit to the works of \citep{simchowitz2020improper}. Let $T$ be a positive integer, interpreted as the horizon of the problem. The initial condition of agent $i$'s state, $i \in [N]$, is $x^i_0 = \omega^{0,i} + \bomega^0$, where $\omega^{0,i} \sim \Ns(0,\Sigma^0)$ and $\bomega^0 \sim \Ns(0,\bSigma^0)$ are i.i.d. noises. For $t \in \{0,\ldots,T-1\}$,
\begin{align} \label{eq:RMS_dyn}
    x^i_{t+1} = A_t x^i_t + \bA_t \bx_t + B_t u^{1,i}_t + \bB_t \bu^1_t + u^{2,i}_t + \bu^2_t + \omega^i_t + \bomega_t, \forall i \in [N]
\end{align}
where $u^{1,i}_t$ is the control action of the $i^{th}$ agent, $\bx_t := \sum_{i = 1}^N x^i_t/N$ is referred to as the state mean-field  and $\bu^j_t := \sum_{i=1}^N u^{j,i}/N$ for $j \in \{1,2\}$ are the control and noise mean-fields respectively. Each agent's dynamics are perturbed by two types of noise: $\omega^i_t$ and $\bomega_t$ are referred to as stochastic noises since they are i.i.d. and their distributions are known ($\omega^i_t \sim \Ns(0,\Sigma)$ and $\bomega_t \sim \Ns(0,\bSigma)$). All of our results (excluding the finite-sample analysis of the RL Algorithm) can be readily generalized for zero-mean non-Gaussian disturbances with finite variance. 

In order to define the robust control problem we define the \emph{performance index} of the population 
which penalizes the deviation of the agents from their (state and control) mean-fields and also regulates the mean-fields: 
\begin{align}
    \hspace{-0.2cm}J_N (\mcu^1, \mcu^2) = \frac{1}{N} \sum_{i=1}^N \EE \sum_{t=0}^{T-1}\Big[ \lVert x^i_t - \bx_t \rVert^2_{Q_t} + \lVert \bx_t \rVert^2_{\bQ_t} + \lVert u^{1,i}_t - \bu^1_t \rVert^2 + \lVert \bu^1_t \rVert^2 \Big] + \lVert x^i_T - \bx_T \rVert^2_{Q_T} + \lVert \bx_T \rVert^2_{\bQ_T} \label{eq:agent_noise}
\end{align}
where the matrices $Q_t,\bQ_t > 0$ are symmetric matrices, $\mcu^j = (\mcu^{j,i})_{i \in [N]}$  
where each $\mcu^{j,i}$ for $j \in \{1,2\}$ is adapted to the distribution information structure i.e. $\sigma$-algebra generated by $x^i_t$ and $\bx_t$ and $\Us^1,\Us^2$ represent the set of all possible $\mcu^1,\mcu^2$, respectively. We define the \emph{noise index} of the population in a similar manner 
\begin{align}
    \varpi_N (\mcu^1, \mcu^2) = \frac{1}{N} \sum_{i=1}^N \EE \sum_{t=0}^{T-1}\Big[ \lVert u^{2,i}_t - \bu^2_t \rVert^2 + \lVert \bu^2_t \rVert^2 + \lVert \omega^i_t \rVert^2 + \lVert \bomega_t \rVert^2 \Big]. \label{eq:varpi_t}
\end{align}
The robust control problem for this $N$ agent system is that of finding the range of noise attenuation levels $\gamma > 0$ such that: 
\begin{align} \label{eq:robust_N_agent_control}
     \exists \mcu^1 \in \Us^1, \forall \mcu^2 \in \Us^2, \qquad J_N (\mcu^1, \mcu^2) \leq \gamma^2  \varpi_N (\mcu^1, \mcu^2)
\end{align}
Any $\gamma$ for which the above inequality is satisfied is referred to as a \emph{viable attenuation level} and the least among them is called the \emph{minimum attenuation level}. The controller $\mcu^1$ which ensures a particular level $\gamma$ of noise attenuation is referred to as the \emph{robust controller} corresponding to $\gamma$ (or robust controller in short). Since the inequality \eqref{eq:robust_N_agent_control} can also be reformulated as $J_N(\cdot)/\varpi_N(\cdot) \leq \gamma^2$, a viable attenuation parameter $\gamma^2$ is also an upper bound on the noise-to-output gain of the system. As outlined in \citep{bacsar2008h} for a single agent problem the condition \eqref{eq:robust_N_agent_control} is equivalent to finding the range of value of $\gamma > 0$ such that
\begin{align}
    \inf_{\mcu^1} \sup_{\mcu^2} \big( J_N (\mcu^1, \mcu^2) - \gamma^2 \varpi_N (\mcu^1, \mcu^2) \big) \leq 0 \label{eq:inf_sup},
\end{align}
where the infimizing controller $u^1$ is the robust controller and the supremizing controller $u^2$ is the worst-case non-stochastic noise. If we define the robust $N$ agent cost $J^\gamma_N$ as follows
\begin{align*}
    J^\gamma_N (\mcu^1, \mcu^2) = & J_N(\mcu^1,\mcu^2) - \gamma^2 \EE \frac{1}{N} \sum_{i=1}^N \sum_{t=0}^{T-1} \big( \lVert u^{2,i}_t - \bu^2_t \rVert^2 + \lVert \bu^2_t \rVert^2 \big),
\end{align*}
then using \eqref{eq:agent_noise} and \eqref{eq:varpi_t}, the robust $N$ agent control problem \eqref{eq:inf_sup} can be equivalently written as
\begin{align}
    \inf_{u^1} \sup_{u^2} J^\gamma_N (\mcu^1,\mcu^2) - \gamma^2 \EE \frac{1}{N} \sum_{i=1}^N \sum_{t=0}^{T-1} (\lVert \omega^i_t \rVert^2 + \lVert \bomega_t \rVert^2)  \leq 0. \label{eq:N_agent_inf_sup}
\end{align}
Due to the distributed information structure of the agents the standard theory of single-agent robust control does not apply in this setting. Hence we are unable to provide sufficient conditions for a given $\gamma > 0$ to be a viable attenuation level, and we resort to the mean-field limit as $N\to\infty$, which is of independent interest. The next section formulates the Robust Mean-Field Control (RMFC) problem 
and its equivalent 2-player zero-sum Mean-Field Type Game (ZS-MFTG) representation, and provides sufficient conditions for solvability of both.
\section{Robust Mean-Field Control} \label{sec:RMFC}
Consider a system with infinitely many agents, where the generic agent has linear dynamics of its state $x_t$ for a finite-horizon $t \in \{0,\ldots,T-1\}$:
\begin{align} \label{eq:MFTG_dyn}
    x_{t+1} = A_t x_t + \bA_t \bx_t + B_t u^{1}_t + \bB_t \bu^1_t + u^{2}_t + \bu^2_t + \omega_t + \bomega_t,
\end{align}
where $u^{1}_t$ is the control action of the generic agent, $\bx_t := \EE[x_t | (\bomega_s)_{0 \leq s \leq t-1}]$ is referred to as the state mean-field and $\bu^j_t := \EE[u^j_t | (\bomega_s)_{0 \leq s \leq t-1}]$ for $j \in \{1,2\}$ are the control and noise mean-fields respectively. The initial condition of the generic agent is $x_0 = \omega^{0} + \bomega^0$, where $\omega^{0} \sim \Ns(0,\Sigma^0)$ and $\bomega^0 \sim \Ns(0,\bSigma^0)$ are i.i.d. noises. The stochastic noises $\omega^i_t$ and $\bomega_t$ are i.i.d. such that 
$\omega^i_t \sim \Ns(0,\Sigma)$ and $\bomega_t \sim \Ns(0,\bSigma)$, whereas the non-stochastic noise $u^{2}_t$ are un-modeled uncertainties. Similar to the $N$ agent case, we define the robust mean-field cost $J^\gamma$ as follows
\begin{align} \label{eq:MFTG_cost}
    J^\gamma (\mcu^1, \mcu^2) = & \EE \sum_{t=0}^{T-1}\Big[ \lVert x_t - \bx_t \rVert^2_{Q_t} + \lVert \bx_t \rVert^2_{\bQ_t} + \lVert u^{1}_t - \bu^1_t \rVert^2 + \lVert \bu^1_t \rVert^2 - \gamma^2 \big( \lVert u^{2}_t - \bu^2_t \rVert^2 + \lVert \bu^2_t \rVert^2 \big) \\
    & \hspace{10.cm} + \lVert x_T - \bx_T \rVert^2_{Q_T} + \lVert \bx_T \rVert^2_{\bQ_T} \Big] .\nonumber
\end{align}
Now the robust mean-field control problem  
which is the mean-field analog to \eqref{eq:N_agent_inf_sup} is defined as follows.
\begin{definition}[Robust Mean-Field Control problem]
If for a given $\gamma > 0$ the following inequality is satisfied, then $\gamma$ is a viable noise attenuation level for the robust mean-field control problem.
    \begin{align}
    \inf_{\mcu^1} \sup_{\mcu^2} J^\gamma (\mcu^1,\mcu^2) - \gamma^2 \EE \sum_{t=0}^{T-1} \lVert \omega_t \rVert^2 + \lVert \bomega_t \rVert^2  \leq 0 .\label{eq:MF_inf_sup}
\end{align}
Moreover, the infimizing controller $\mcu^1$ in \eqref{eq:MF_inf_sup} is a robust controller (corresponding to $\gamma$). 
\end{definition}

Now, under the condition of interchangability of the inf and sup operations, the problem of finding \newline $\inf_{\mcu^1} \sup_{\mcu^2} J^\gamma (\mcu^1,\mcu^2)$ is that of finding the Nash equilibrium (equivalently, saddle point, in this case) of the \emph{Zero-sum 2-player Mean-Field Type Game}; see \citep{carmona2020policy,carmona2021linear} for a very similar LQ setting without  
the theoretical analysis of the RL algorithm. 
In the following section we provide sufficient conditions for existence and uniqueness of a solution to this saddle point problem along with the value of $\inf_{\mcu^1} \sup_{\mcu^2} J^\gamma (\mcu^1,\mcu^2)$. 

\noindent\textbf{$2$-player Zero-sum Mean-Field Type Games: }
Let us define $y_t = x_t - \bx_t, z_t = \bx_t$. The dynamics of $y_t$ and $z_t$ can be written as
\begin{align*}
	y_{t+1} = A_t y_t + B_t(u^1_t - \bu^1_t) + u^2_t - \bu^2_t + \omega_t - \bomega_t, \hspace{0.2cm}
	z_{t+1} = \tA_t z_t + \tB_t \bu^1_t + 2 \bu^2_t + 2 \bomega_t,
\end{align*}
where $\tA_t = A_t + \bA_t$ and $\tB_t = B_t + \bB_t$. The optimal controls are known to be linear \citep{carmona2020policy}, hence we restrict our attention the set of linear controls in $y_t$ and $z_t$,
\begin{align*}
	u^1_t = \mcu^1_t(x_t,\bx_t) = -K^1_t(x_t - \bx_t) - L^1_t \bx_t , \hspace{0.2cm}
	u^2_t = \mcu^2_t(x_t,\bx_t) = K^2_t(x_t - \bx_t) + L^2_t \bx_t
\end{align*}
which implies that $\bu^1_t = -L^1_t \bx_t$ and $\bu^2_t = L^2_t \bx_t$. The dynamics of the processes $y_t$ and $z_t$ can be re-written as
\begin{align}
	y_{t+1} = (A_t- B_t K^1_t + K^2_t) y_t + \omega_t - \bomega_t,  \hspace{0.2cm}
	z_{t+1} = (\tA_t - \tB_t L^1_t +   L^2_t) z_t + 2\bomega_t. \label{eq:y_z_dyn}
\end{align}
Since the dynamics of $y_t$ and $z_t$ are decoupled, we can decompose the cost $J^\gamma$ into the following two parts:
\begin{align}
	J^\gamma(K,L) & = J^\gamma_y(K) + J^\gamma_z(L), \nonumber \\
	J^\gamma_y(K) & = \EE \Big[ \sum_{t=0}^{T-1}   y_t^\top(Q_t + (K^1_t)^\top K^1_t - \gamma^2 (K^2_t)^\top  K^2_t) y_t  + y_T^\top Q_T y_T\Big], \label{eq:J_gamma_decop} \\
	J^\gamma_z(L) & =  \EE \Big[ \sum_{t=0}^{T-1}   z_t^\top(\bQ_t + (L^1_t)^\top L^1_t - \gamma^2 (L^2_t)^\top  L^2_t) z_t + z^\top_T \bQ_T z_T\Big]. \nonumber
\end{align}
The 2-player MFTG \eqref{eq:MFTG_dyn}-\eqref{eq:MFTG_cost} has been decoupled into two 2-player LQ dynamic game problems as shown below: 
\begin{align*}
	\min_{K^1,L^1} \max_{K^2,L^2} J^\gamma((K^1,K^2),(L^1,L^2)) = \min_{K^1} \max_{K^2} J^\gamma_y(K) + \min_{L^1} \max_{L^2} J^\gamma_z(L)
\end{align*}
where the dynamics of $y_t$ and $z_t$ are defined in \eqref{eq:y_z_dyn}. In the following section, using results in the literature, we specify the sufficient conditions for existence and uniqueness of Nash equilibrium of the 2-player MFTG and also present the \emph{value} (Nash cost) of the game. Building on the techniques developed in \citep{bacsar1998dynamic,carmona2020policy}, we can prove the following result.
\begin{theorem} \label{thm:NE_2P_MFTG}
    Assume for a given $\gamma > 0$,
    \begin{align} \label{eq:gamma_cond_1}
        \gamma^2 I - M^\gamma_t > 0 \text{ and } \gamma^2 I - \bM^\gamma_t > 0,
    \end{align}
    where $M^\gamma_t$ and $\bM^\gamma_t$ are positive semi-definite matrices which satisfy the Coupled Algebraic Riccati equations,
    \begin{align}
        M^\gamma_t = Q_t + A^\top_t M^\gamma_{t+1} \Lambda^{-1}_t A_t, \hspace{0.2cm} \Lambda_t = I + (B_t B^\top_t - \gamma^{-2} I) M^\gamma_{t+1}, \hspace{0.2cm} M^\gamma_T = Q_T, \nonumber \\
        \bM^\gamma_t = \bQ_t + \tA^\top_t \bM^\gamma_{t+1} \bLambda^{-1}_t \tA_t ,\hspace{0.2cm} \bLambda_t = I + (\tB_t \tB^\top_t - \gamma^{-2} I) \bM^\gamma_{t+1}, \hspace{0.2cm} \bM^\gamma_T = \bQ_T \label{eq:M_t_bM_t} \\
        N^\gamma_t = N^\gamma_{t+1} + \tr(M^\gamma_{t+1} \Sigma), \hspace{0.2cm} N^\gamma_T = 0, \hspace{0.2cm} \bN^\gamma_t = \bN^\gamma_{t+1} + \tr(\bM^\gamma_{t+1} \Sigma), \hspace{0.2cm} \bN^\gamma_T = 0. \nonumber
    \end{align}
    Then, $u^{1*}_t = - K^{1*}_t (x_t - \bx_t) - L^{1*}_t \bx_t$ and $u^{2*}_t = K^{2*}_t (x_t - \bx_t) + L^{2*}_t \bx_t$ (complete expressions provided in Supplementary Materials) are the unique Nash policies. Furthermore, the Nash equilibrium (equivalently, saddle point) value is
    \begin{align}
        \inf_{\mcu^1} \sup_{\mcu^2} J^\gamma (\mcu^1,\mcu^2) = \tr( M^\gamma_0 \Sigma^0) + \tr( \bM^\gamma_0 \bSigma^0) + N^\gamma_0 + \bN^\gamma_0 \label{eq:Nash_value}
    \end{align}
\end{theorem}
This result can be proved using techniques in proofs of Theorem 3.2 in \citep{bacsar2008h} or Proposition 36 in \citep{carmona2021linear}. 
We now use the Nash value of the game \eqref{eq:Nash_value} to come up with a condition for the attenuation level $\gamma $ which solves the robust mean-field control problem \eqref{eq:MF_inf_sup}. First we simplify expression in \eqref{eq:MF_inf_sup} $\EE \sum_{t=0}^{T-1} \lVert \omega_t \rVert^2 + \lVert \bomega_t \rVert^2 = T \tr(\Sigma + \bSigma)$ using the i.i.d. stochastic nature of the noise. Combining this fact with \eqref{eq:Nash_value}, we arrive at the conclusion that \eqref{eq:MF_inf_sup} will be satisfied if and only if
\begin{align} \label{eq:gamma_cond_2}
    \sum_{t=1}^{T} \tr((M^\gamma_t - \gamma^2 I) \Sigma + (\bM^\gamma_t - \gamma^2 I )\bSigma) +\tr( M^\gamma_0 \Sigma^0) + \tr( \bM^\gamma_0 \bSigma^0) \leq 0
\end{align}

\noindent Notice that the conditions \eqref{eq:gamma_cond_1} and \eqref{eq:gamma_cond_2} are different, as the first one requires positive definiteness of matrices and the second one requires a scalar inequality. 
Now we solve the robust $N$ agent control problem by providing sufficient conditions for a given attenuation level $\gamma$ satisfying \eqref{eq:robust_N_agent_control}. 
\begin{theorem} \label{thm:MFTG_to_RMC}
    Let $\gamma > 0$. Assume, in addition to \eqref{eq:gamma_cond_1}, that we also have 
    \begin{align} \label{eq:robust_MFTG_MFC}
    \sum_{t=1}^{T} \tr((M^\gamma_t - \gamma^2 I) \Sigma + (\bM^\gamma_t - \gamma^2 I )\bSigma) +\tr( M^\gamma_0 \Sigma^0) + \tr( \bM^\gamma_0 \bSigma^0) \leq -\frac{CT}{N},
\end{align}
where $C$ is a constant which depends only on the model parameters and $M^\gamma_t$ and $\bar M^\gamma_t$ \eqref{eq:M_t_bM_t}. Then $\gamma$ is a viable attenuation level for the Robust $N$ agent control problem \eqref{eq:robust_N_agent_control}. Moreover the robust controller for each agent $i$ is given by $u^{1,i*}_t = - K^{1*}_t (x^i_t - \bx_t) - L^{1*}_t \bx_t$.
\end{theorem}
The proof of this result can be found in the full version of this paper \citep{zaman2024robust}. 
The above theorem states that, if for a given $\gamma$, conditions \eqref{eq:gamma_cond_1} and \eqref{eq:robust_MFTG_MFC} are satisfied (given that $M^\gamma_t$ and $\bM^\gamma_t$ are defined by \eqref{eq:M_t_bM_t}), then not only is $\gamma$ a viable attenuation level for the original Robust multi-agent control problem \eqref{eq:RMS_dyn}-\eqref{eq:robust_N_agent_control}, but the Nash equilibrium for the ZS-MFTG also yields the robust controller $u^{1,i*}_t = - K^{1*}_t (x^i_t - \bx_t) - L^{1*}_t \bx_t$ for the original finite-agent game. Condition \eqref{eq:robust_MFTG_MFC} is strictly stronger than condition \eqref{eq:gamma_cond_2} but approaches \eqref{eq:robust_MFTG_MFC} as $N \rightarrow \infty$.
\section{Reinforcement Learning for Robust Mean-Field Control} \label{sec:RL_MFTG} 
In this section we present the Receding-horizon policy Gradient Descent Ascent (RGDA) algorithm to compute the Nash equilibrium (Theorem \ref{thm:NE_2P_MFTG}) of the 2-player MFTG \eqref{eq:MFTG_dyn}-\eqref{eq:MFTG_cost}, which will also generate the robust controller for a fixed noise attenuation level $\gamma$. For this section we assume access to only the finite-horizon costs of the agents under a set of control policies, and not the state trajectories. Under this setting the model of the agents cannot be constructed hence our approach is \emph{truly} model free \citep{malik2019derivative}. 
Due to the non-convex non-concave (also non-coercive \citep{zhang2020arobust}) nature of the cost function $J^\gamma$ in \eqref{eq:J_gamma_decop}, 
instead we solve the receding-horizon problem, for each $t = \{T-1,\ldots,1,0\}$ backwards-in-time. This entails solving $2 \times T$ min-max problems, where each problem is convex-concave and aims at finding $(K_t,L_t) = \big( (K^1_t,K^2_t), (L^1_t,L^2_t) \big)$ at time step $t$, given the set of \emph{future} controllers (controllers for times greater than $t$), $\big( (\tK_{t+1},\tL_{t+1}), \ldots, (\tK_T,\tL_T) \big)$ are held constant. But first we must approximate the mean-field term using finitely many agents. 

\noindent\textbf{Approximation of mean-field terms using $M$ agents:} Since simulating infinitely many agents is impractical, in this section we outline how to use a set of $2 \leq M < \infty$ agents to approximately simulate the mean-field in a MFTG. Each of the $M$ agents has state $x^i_t$ at time $t$ where $i \in [M]$. The agents follow controllers linear in their private state and empirical mean-field, $x^i_t$ and $\tz_t$, respectively, 
	$u^1_t = -K^1_t (x^i_t - \tz_t) - L^1_t \tz_t, 
 u^2_t = K^2_t (x^i_t - \tz_t) + L^2_t \tz_t,$
where the empirical mean-field is $\tz_t := \frac{1}{M} \sum_{i = 1}^M x^i_t$. Under these control laws, the dynamics of agent $i \in [M]$ are
\begin{align*}
	x^i_{t+1} = (A_t- B_t K^1_t +  K^2_t) (x^i_t - \tz_t) + (\tA_t - \tB_t L^1_t +  L^2_t) \tz_t + \omega^i_{t+1} + \bomega_t
\end{align*}
and the dynamics of the empirical mean-field $\tz_t$ is
	$\tz_{t+1} = (\tA_t - \tB_t L^1_t + L^2_t) \tz_t + \tomega^0_{t+1}, \hspace{0.2cm} $ where $ 
 \tomega^0_{t+1} =  \bomega_t + \frac{1}{M} \sum_{i=1}^M \omega^i_{t+1}.$
The cost of each agent is
\begin{align*}
	\tJ^{i,\gamma}(u_1, u_2) = & \EE \Big[\sum_{t=0}^{T-1}   (x^i_t - \tz_t)^\top [Q_t + (K^1_t)^\top K^1_t - \gamma^2 (K^2_t)^\top  K^2_t] (x^i_t - \tz_t) + (x^i_T - \tz_T)^\top Q_T (x^i_T - \tz_T) \\
    & \hspace{6cm} + \tz_t^\top [\bQ_t + (L^1_t)^\top  L^1_t - \gamma^2 (L^2_t)^\top L^2_t] \tz_t + \tz_T^\top \bQ_T \tz_T \Big].
\end{align*}
Now, similarly to the previous section, we define $y^i_t = x^i_t - \tz_t$. 
The dynamics of $y^i_t$ are
	$y^i_{t+1} = (A_t- B_t K^1_t + K^2_t) y^i_t + \tomega^i_{t+1},
 $ where $\tomega^i_{t+1} = \frac{M-1}{M} \omega^i_{t+1} - \frac{1}{M} \sum_{j \neq i} \omega^j_{t+1}$. 
The cost can then be decomposed in a manner similar to \eqref{eq:J_gamma_decop}:
\begin{align}
	\tJ^{i,\gamma} \big((K^1_t,K^2_t),(L^1_t,L^2_t) \big) & = \tJ^{i,\gamma}_y(K^1_t,K^2_t) + \tJ^{i,\gamma}_z(L^1_t,L^2_t), \nonumber \\
	\tJ^{i,\gamma}_y(K^1_t,K^2_t) & = \EE \Big[\sum_{t=0}^{T-1}   (y^i_t)^\top [Q_t + (K^1_t)^\top K^1_t - \gamma^2 (K^2_t)^\top K^2_t] y^i_t + (y^i_T)^\top Q_T y^i_T\Big],  \label{eq:tJ} \\
	\tJ^{i,\gamma}_z(L^1_t,L^2_t) & = \EE \Big[\sum_{t=0}^{T-1}   \tz_t^\top [\bQ_t + (L^1_t)^\top  L^1_t - \gamma^2 (L^2_t)^\top  L^2_t] \tz_t + \tz_T^\top \bQ_T \tz_T \Big]. \nonumber 
\end{align}
\noindent\textbf{Receding-horizon approach:} Similar to the approach in Section \ref{sec:form}, instead of finding the optimal, $K^*$ and $L^*$ which optimizes $\tJ$ in \eqref{eq:tJ}, we solve the receding-horizon problem for each $t = \{T-1,,\ldots,1,0\}$ backwards-in-time. 
This forms two decoupled min-max convex-concave problems of finding $(K_t,L_t) = \big( (K^1_t,K^2_t), (L^1_t,L^2_t) \big)$ at each time step $t$, given the set of controllers for times greater than $t$, $\big( (\tK_{t+1},\tL_{t+1}), \ldots, (\tK_T,\tL_T) \big)$
\begin{align}
	& \min_{(K^1_t,L^1_t)} \max_{(K^2_t,L^2_t)} \tJ^{i,\gamma}_t(K_t,L_t) = \nonumber \\
	& \underbrace{\EE \Big[   y_t^\top (Q_t + (K^1_t)^\top K^1_t - \gamma^2 (K^2_t)^\top K^2_t ) y_t + \sum_{k=t+1}^T  y_k^\top (Q_t + (\tK^1_k)^\top  \tK^1_k - \gamma^2 (\tK^2_k)^\top \tK^2_k ) y_k \Big]}_{\tJ^{i,\gamma}_{y,t}} \label{eq:reced_hor_cost}\\
	& +  \underbrace{\EE \Big[   z_t^\top (\bQ_t + (L^1_t)^\top  L^1_t - \gamma^2 (L^2_t)^\top L^2_t ) z_t + \sum_{k=t+1}^T  z_k^\top (\bQ_t + \tL^\top_{1,k}  \tL^1_k - \gamma^2 (\tL^2_k)^\top \tL^2_k ) z_k \Big]}_{\tJ^{i,\gamma}_{z,t}}, \nonumber 
\end{align}
for any $i \in [M]$ and $y_t \sim \Ns(0, \Sigma_y), z_t \sim \Ns (0, \Sigma_z)$. This receding-horizon problem is solved using Receding-horizon policy Gradient Descent Ascent (RGDA) (Algorithm \ref{alg:RL^2_tP_MFTG}) where at each time instant $t$ the Nash control is approached using gradient descent ascent. We anticipate a small approximation error between the optimal controller and its computed approximation $\tK_t$ (respectively $\tL_t$). However, this error is shown to be well-behaved (Theorem \ref{thm:main_res}), as we progress backwards-in-time, given that the hyper-parameters of RGDA satisfy certain bounds.

\noindent \textbf{Receding-horizon policy Gradient Descent Ascent (RGDA) Algorithm:} The RGDA Algorithm (Algorithm \ref{alg:RL^2_tP_MFTG} is a bi-level optimization algorithm where the outer loop starts at time $t=T-1$ and moves backwards-in-time, and the inner loop is a gradient descent (for control parameters $(K^1_t, L^1_t)$) ascent (for control policy $(K^2_t, L^2_t)$) update with learning rate $\eta_k$. The gradient descent ascent step entails computing an approximation of the \emph{exact} gradients of cost $\tJ^{i,\gamma}_t$ with respect to the controls variables $(K^1_t, L^1_t),(K^2_t, L^2_t)$. To obtain this approximation in a data driven manner we utilize a zero-order stochastic gradient $\tnabla_1 \tJ^{i,\gamma}_t(K_t,L_t), \tnabla_2 \tJ^{i,\gamma}_t(K_t,L_t)$ \citep{fazel2018global,malik2019derivative} which requires cost computation under a given set of controllers \eqref{eq:reced_hor_cost} as shown below.
\begin{align*}
    \tnabla_1 \tJ^{i,\gamma}_t(K_t,L_t) & = \frac{n}{Mr^2} \sum_{j=1}^M  \tJ^{i,\gamma}_t((K^{j,1}_t,K^2_t),(L^{j,1}_t,L^2_t)) e_j, \hspace{0.2cm} \begin{pmatrix} K^{j,1}_t \\ L^{j,1}_t \end{pmatrix} = \begin{pmatrix} K^1_t \\ L^1_t \end{pmatrix} + e_j, \hspace{0.2cm} e_j \sim \mathbb{S}^{n-1}(r) \\
    \tnabla_2 \tJ^{i,\gamma}_t(K_t,L_t) & = \frac{n}{Mr^2} \sum_{j=1}^M  \tJ^{i,\gamma}_t((K^1_t,K^{j,2}_t),(L^1_t,L^{j,2}_t)) e_j, \hspace{0.2cm} \begin{pmatrix} K^{j,2}_t \\ L^{j,2}_t \end{pmatrix} = \begin{pmatrix} K^2_t \\ L^2_t \end{pmatrix} + e_j, \hspace{0.2cm} e_j \sim \mathbb{S}^{n-1}(r).
\end{align*}
Stochastic gradient computation entails computing the cost of $N_b$ different \emph{perturbed} controllers, with a perturbation magnitude if $r$ also called the \emph{smoothing radius}. This stochastic gradient provides us with a \emph{biased} approximation of the exact gradient whose \emph{bias} and \emph{variance} can be controlled by tuning the values of $N_b$ and $r$. Finally to ensure stability of the learning algorithm, we use projection $\proj_D$ onto a $D$-ball such that the norm of the matrices is bounded by $D$, $\lVert (K_t,L_t) \rVert^2 \leq D$. The radius of the ball $D$ is chosen such that the Nash equilibrium controllers lie within this ball.

\begin{algorithm}[h!]
	\caption{RGDA Algorithm for 2-player MFTG}
	\begin{algorithmic}[1] \label{alg:RL^2_tP_MFTG}
		\FOR {$t = T-1,\ldots,1,0,$}
			\STATE {Initialize $K_t = (K^1_t,K^2_t)=0, L_t = (L^1_t,L^2_t) = 0$}
			\FOR {$k = 0,\ldots,K$}
				\STATE {\underline{\bf Gradient Descent}} 
					$\begin{pmatrix} K^1_t \\ L^1_t	\end{pmatrix} \leftarrow \proj_D \bigg( \begin{pmatrix} K^1_t \\ L^1_t	\end{pmatrix} - \eta_k \tnabla_1 \tJ^{i,\gamma}_t(K_t,L_t) \bigg), $
			\STATE {\underline{\bf Gradient Ascent}}
				$\begin{pmatrix} K^2_t \\ L^2_t	\end{pmatrix} \leftarrow \proj_D \bigg( \begin{pmatrix} K^2_t \\ L^2_t	\end{pmatrix} + \eta_k \tnabla_2 \tJ^{i,\gamma}_t(K_t,L_t) \bigg),$ 
			\ENDFOR
		\ENDFOR
	\end{algorithmic}
\end{algorithm}

\noindent\textbf{RGDA algorithm analysis: } In this section we start by showing linear convergence of the inner loop gradient descent ascent (Theorem \ref{thm:inner_loop_conv}), which is made possible by the convex-concave property of the cost function under the receding horizon approach \eqref{eq:reced_hor_cost}. Then we show that if the error accumulated in each inner loop computation is small enough, the total accumulated error is well behaved (Theorem \ref{thm:main_res}). 

We first define some relevant notation. We define the \emph{joint controllers} for each timestep $t$ as $\bK_t = [(K^1_t)^\top, (K^2_t)^\top]^\top$ and $\bL_t = [(L^1_t)^\top,(L^2_t)^\top]^\top$, for the sake of conciseness. For each timestep $t \in \{T-1,\ldots,1,0\}$ let us also define the \emph{target} joint controllers $\tbK^*_t = (\tK^{1*}_t,\tK^{2*}_t),\tbL^*_t = (\tL^{1*}_t,\tL^{2*}_t)$, as the set of policies which exactly solve the receding-horizon min-max problem \eqref{eq:reced_hor_cost}. Notice that the set of target controllers $\tbK^*_t,\tbL^*_t$ are unique (due to convex-concave nature of \eqref{eq:reced_hor_cost}) but do depend on the set of future joint controllers $(\bK_s,\bL_s)_{t < s < T}$. On the other hand, the Nash joint controllers are denoted by $\bK^*_t = (K^{1*}_t,K^{2*}_t)$ and $\bL^*_t = (L^{1*}_t,L^{2*}_t)$. Furthermore, the target joint controllers are equal to the Nash joint controllers $(\tbK^*_t,\tbL^*_t) = (\bK^*_t,\bL^*_t)$ only if the future joint controllers are also Nash $(\bK_s,\bL_s)_{t < s < T} = (\bK^*_s,\bL^*_s)_{t < s < T}$.

\begin{theorem} \label{thm:inner_loop_conv}
    If the learning rate $\eta_k$ is smaller than a certain function of model parameters, the number of inner loop iterations $K = \Omega(\log(1/\epsilon)),$ the mini-batch size $N_b = \Omega(1/\epsilon)$ and the smoothing radius $r = \Os(\epsilon)$, then at each timestep $t \in \{T-1,\ldots,1,0\}$ the optimality gaps are $\lVert \bK_t - \tbK^*_t \rVert^2_2 \leq \epsilon$ and $\lVert \bL_t - \tbL^*_t \rVert^2_2 \leq \epsilon$.
\end{theorem}
Closed form expressions of the bounds can be found in the proof given in the full version of the paper \citep{zaman2024robust}. 
The linear rate of convergence is made possible by building upon the convergence analysis of descent ascent in \citep{fallah2020optimal} due to the convex-concave nature of the cost function \eqref{eq:reced_hor_cost}. The proof generalizes the techniques used in \citep{fallah2020optimal} to stochastic unbiased gradients by utilizing the fact that the bias in stochastic gradients $\tnabla_j \tJ^{i,\gamma}_t$ for $j \in \{1,2\}$ can be reduced by reducing the smoothing radius $r$. This in turn causes an increase in the variance of the stochastic gradient  which is controlled by increasing the mini-batch size $N_b$.


Now we present the non-asymptotic convergence guarantee of the paper stating that even though each iteration of the outer loop (as timestep $t$ moves backwards-in-time) accumulates error, if the error in each outer loop iteration is small enough, the total accumulated error will also be small enough. The proof can be found in the complete version of the paper \citep{zaman2024robust}. 
\begin{theorem} \label{thm:main_res}
    If all conditions in Theorem \ref{thm:inner_loop_conv} are satisfied, then $\max_{j \in \{1,2\}} \lVert K^j_t - K^{j*}_t \rVert = \Os(\epsilon)$ and $\max_{j \in \{1,2\}} \lVert L^j_t - L^{j*}_t \rVert = \Os(\epsilon)$ for a small $\epsilon > 0$ and $t \in \{T-1,\ldots,0\}$. 
\end{theorem}
The Nash gaps at each time $t$, $\lVert K^j_t - K^{j*}_t \rVert$ and $\lVert L^j_t - L^{j*}_t \rVert$ for $j \in \{1,2\}$ are due to a combination of the optimality gap in the inner loop $\lVert \bK_t - \tbK^*_t \rVert^2_2, \lVert \bL_t - \tbL^*_t \rVert^2_2$ and the accumulated Nash gap in the future joint controllers $\lVert K^j_s - K^{j*}_s \rVert$ and $\lVert L^j_s - L^{j*}_s \rVert$ for $j \in \{1,2\}$ and $t < s < T$. The proof of Theorem \ref{thm:main_res} characterizes these two quantities and then shows that if the optimality gap at each timestep $t \in \{0,\ldots, T-1\}$ never exceeds some small $\epsilon$, then the Nash gap at any time $t$ never exceeds $\epsilon$ scaled by a constant. 

\section{Numerical Analysis}

\begin{figure}
    \centering
    \subfigure[]{\includegraphics[width=0.33\textwidth]{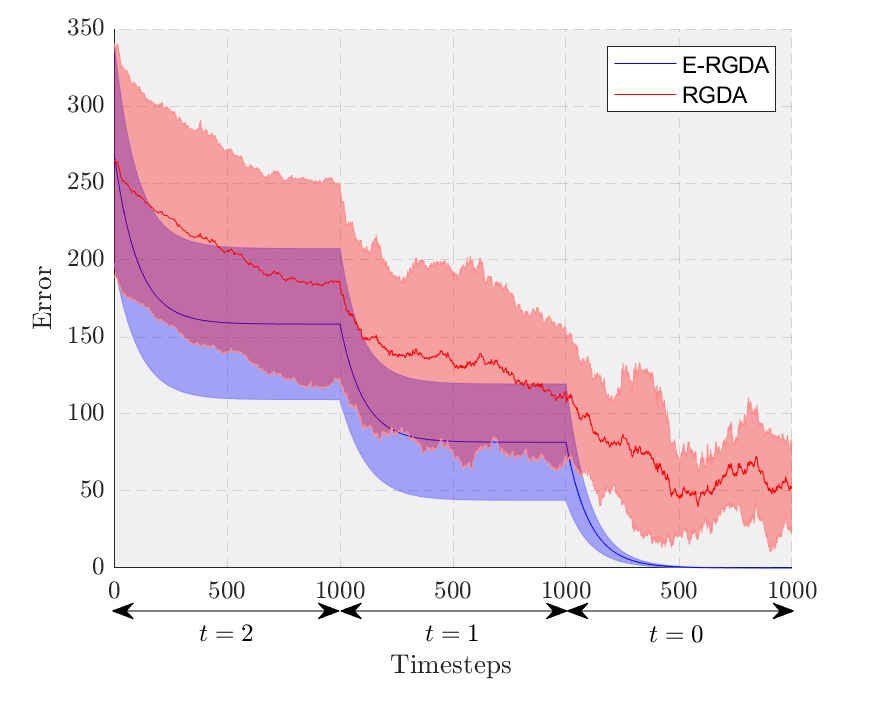} \label{fig:RGDA_vs_ERGDA_1}}
    \subfigure[]{\includegraphics[width=0.33\textwidth]{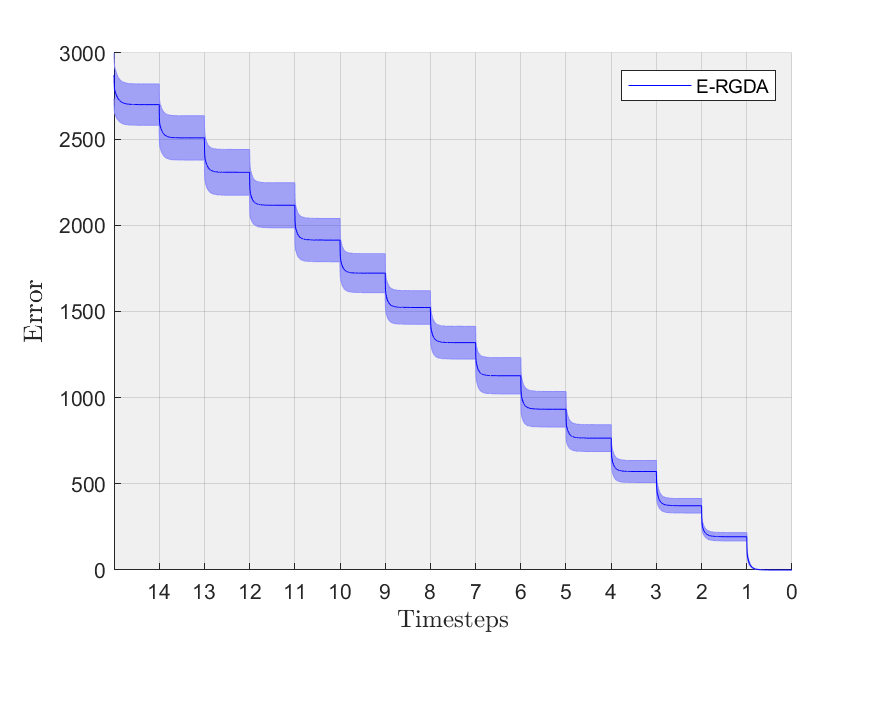}  \label{fig:RGDA_vs_ERGDA_2}}
    \caption{\begin{footnotesize}Performance of RGDA Algorithm\end{footnotesize}}\label{fig:RGDA_vs_ERGDA}
\end{figure}
First, we simulate the RGDA algorithm for time horizon $T=3$, number of agents $M=1000$ and the dimension of the state and action spaces $m=p=2$. For each timestep $t \in \{2,1,0\}$, 
the number of inner-loop iterations $K=1000$, the mini-batch size $N_b = 5 \times 10^4$ and the learning rate $\eta_k = 0.001$. In Figure \ref{fig:RGDA_vs_ERGDA_1} we compare the RGDA algorithm (Algorithm \ref{alg:RL^2_tP_MFTG}) with its exact version (E-RGDA) which has access to the exact policy gradients $\nabla_1 \tJ^{i,\gamma}_t = \delta \tJ^{i,\gamma}_t/\delta (K^1_t, L^1_t )$ and $\nabla_2 \tJ^{i,\gamma}_t = \delta \tJ^{i,\gamma}_t/\delta ( K^2_t, L^2_t )$ at each iteration $k \in [K]$. The error plots in Figures \ref{fig:RGDA_vs_ERGDA_1} and \ref{fig:RGDA_vs_ERGDA_2} show the mean (solid lines) and standard deviation (shaded regions) of error, which is the norm of difference between iterates and Nash controllers. In Figure \ref{fig:RGDA_vs_ERGDA_1} the blue plot shows error convergence of the E-RGDA algorithm, which computes the Nash controllers for the last timestep $t=2$ (using gradient descent ascent with exact gradients) and moves backwards in time. Since at each timestep it has good convergence to Nash policies, the convexity-concavity of cost function at the next timestep is ensured, which results in linear convergence. The red plot in Figure \ref{fig:RGDA_vs_ERGDA_1} shows the error convergence in the RGDA algorithm which uses stochastic gradients, which results in a noisy but downward trend in error. Notice that RGDA imitates E-RGDA in a noisy fashion and at each timestep the iterates only approximate the Nash controllers. This approximation can be further sharpened by increasing the mini-batch size $N_b$ and decreasing smoothing radius $r$. Figure \ref{fig:RGDA_vs_ERGDA_1} shows the error convergence of E-RGDA for a ZS-MFTG with  $T=15$ and state and action space dimensions $m=p=2$. 

\begin{figure}
    \centering
    \includegraphics[width=0.75\textwidth]{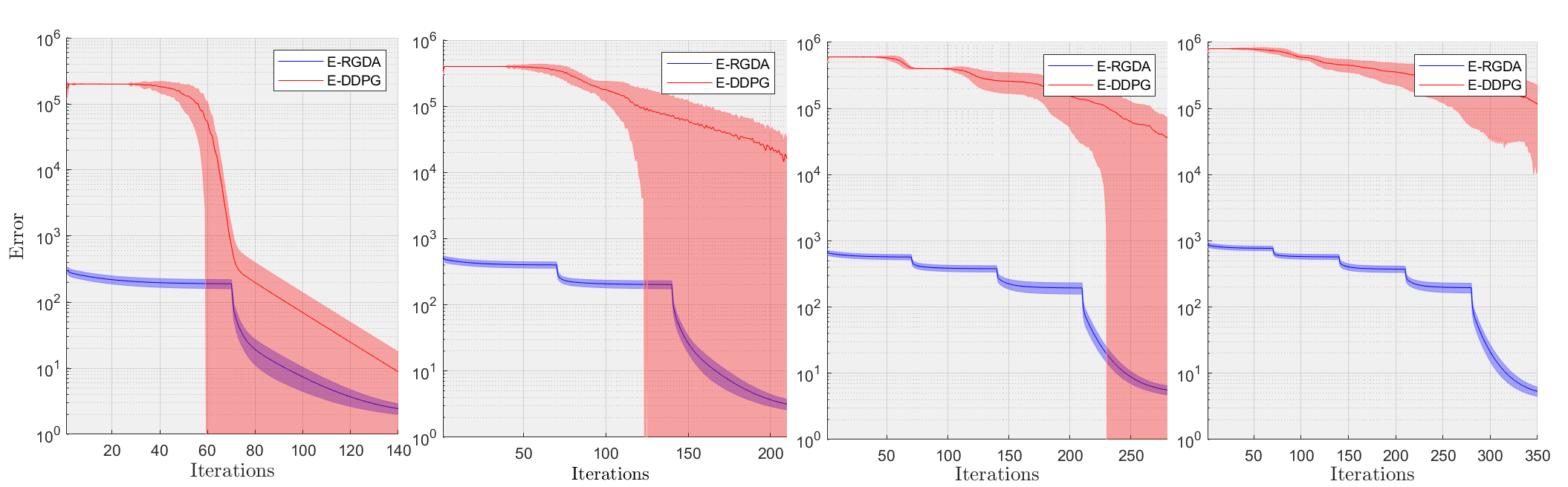} 
    \caption{\begin{small}Comparison between E-RGDA and E-DDPG. The time-horizon is increasing from left to right with $T=2$ (left-most), $T=3$ (center left), $T=4$ (center right) and $T=5$ (right-most)\label{fig:E-RGDA_vs_E-DDPG}\end{small}}
\end{figure}

Figure \ref{fig:E-RGDA_vs_E-DDPG} compares the E-RGDA algorithm with the exact 2-player zero-sum version of the MADPG algorithm (referred to as E-DDPG) \citep{lowe2017multi} 
which serves as a baseline as it does not use the receding-horizon approach. The number of inner-loop iterations for E-RGDA is $K=70$ and the learning rate for both algorithms is $\eta = 0.025$. The four figures represent the comparisons for $T = \{2,3,4,5\}$ and the y-axis is scaled in a logarithmic manner to best show the behavior of the algorithms. For all $T > 1$ the E-DDPG first diverges until it reaches the projection threshold then eventually starts to converge. This is due to the fact that errors in later timesteps cause the convexity-concavity condition to fail resulting in divergence in earlier timesteps. Over time the error decreases in the later timesteps, which causes the error in earlier timesteps to gradually decrease as well. But as seen from Figure~\ref{fig:E-RGDA_vs_E-DDPG}, the convergence for E-DDPG takes significantly longer as the time-horizon increases.

\section{Conclusion}
In this paper, we solve an MARL problem with the objective of designing robust controllers in the presence of modeled and un-modeled uncertainties. We introduce the concept of Robust Mean Field Control (RMFC) problem as the limiting problem when the number of agents grows to infinity. We then establish a connection with Zero-Sum Mean-Field Type Games (ZS-MFTG). 
We resort to the Linear-Quadratic (LQ) structure which, combined with the mean-field approximation, helps to have a more tractable model and to help resolve the analytical difficulty induced by the distributed information structure. This helps us obtain sufficient conditions for robustness of the problem as well as characterization of the robust control policy. We design and provide non-asymptotic analysis of a receding-horizon based RL algorithm which renders the non-coercive cost as convex-concave. Through numerical analysis the receding-horizon approach is shown to ameliorate the overshooting problem observed in the performance of the vanilla algorithm. 
In future work we would like to explore this type of robust mean-field problems beyond the LQ setting and to develop RL algorithms which go beyond the gradient descent-ascent updates used in this paper. Furthermore, our work is a first step in the direction of using mean-field approximations to study robust MARL problems which occur in many real-world scenarios, but the study of concrete examples is left for future work.

\section*{Acknowledgments}
The authors would like to thank Xiangyuan Zhang (University of Illinois Urbana-Champaign) for useful discussions regarding the Receding-horizon Policy Gradient algorithm. 

 \section*{Disclaimer:} This paper was prepared for informational purposes in part by the Artificial Intelligence Research group of JP Morgan Chase \& Co and its affiliates (“JP Morgan”), and is not a product of the Research Department of JP Morgan. JP Morgan makes no representation and warranty whatsoever and disclaims all liability, for the completeness, accuracy or reliability of the information contained herein. This document is not intended as investment research or investment advice, or a recommendation, offer or solicitation for the purchase or sale of any security, financial instrument, financial product or service, or to be used in any way for evaluating the merits of participating in any transaction, and shall not constitute a solicitation under any jurisdiction or to any person, if such solicitation under such jurisdiction or to such person would be unlawful.

\bibliography{references}

\newpage
\input{sec_appendix}

\end{document}

%% file: sec_appendix.tex
\begin{centering}
    \hfill \textbf{\Large Supplementary Materials} \hfill
\end{centering}
\vspace{0.2cm}

\section*{Proof of Theorem \ref{thm:MFTG_to_RMC}}
\begin{proof}
    Central to this analysis is the quantification of the difference between the finite and infinite population costs for a given set of control policies. First we express the state and mean-field processes in terms of the noise processes, for the finite and infinite population settings. This then allows us to write the costs (in both settings) as quadratic functions of the noise process, which simplifies quantification of the difference between these two costs.
    
    Let us first consider the finite agent setting with $M$ number of agents. Consider the dynamics of state $x^j$ of agent $j \in [M]$ under the NE of the MFTG (Theorem \ref{thm:NE_2P_MFTG})
    \begin{align} \label{eq:epsN_x_M}
        x^{j*,M}_{t+1} = (A_t - B_t K^{1*}_t + K^{2*}_t) x^{j*,M}_t + (\bA_t - B_t (L^{1*}_t - K^{1*}_t) - \bB_t L^{1*}_t + L^{2*}_t - K^{2*}_t)\bx^{M*}_t + \omega^j_{t} + \bomega_t
    \end{align}
    where the superscript $M$ denotes the dynamics in the finite population game \eqref{eq:RMS_dyn} and $\bx^{M*} = \frac{1}{M}\sum_{j \in [M]} x^{j*,M}_t$ is the empirical mean-field. We can also write the dynamics of the empirical mean-field as
    \begin{align} \label{eq:epsN_bx_M}
        \bx^{M*}_{t+1} = \underbrace{\big( A_t + \bA_t - B_t L^{1*}_t + L^{2*}_t \big)}_{\tL^*_t}\bx^{M*}_t + \underbrace{\frac{1}{M} \sum_{j \in [M]}\omega^j_{t} + \bomega_t}_{\bomega^M_{t}}.
    \end{align}
    For simplicity we assume that $x^{j*,M}_0 = \omega^j_0 + \bomega_0$ which also implies that $\bx^{M*}_0 = \bomega^M_0$. Using \eqref{eq:epsN_bx_M} we get the recursive definition of $\bx^M_t$ as
    \begin{align*}
        \bx^{M*}_t = \sum_{s=0}^t \tL^*_{[t-1,s]}\bomega^N_s, \text{ where } \tL^*_{[s,t]} := \tL^*_s \tL^*_{s-1} \ldots \tL^*_t \text{, if } s \geq t. \text{ and } \tL^*_{[s,t]} = I \text{ otherwise}.
    \end{align*}
    Hence $\bx^{M*}_t$ can be characterized as a linear function of the noise process
    \begin{align*}
        \bx^{M*}_t = \big( \bPsi^* \bomega^M \big)_t, \text{ where } \bPsi^* = 
        \begin{pmatrix}
            I & 0 & 0 & \hdots \\
            \tL^*_{[0,0]} & I & 0 & \hdots \\
            \tL^*_{[1,0]} & \tL^*_{[1,1]} & I & \hdots \\
            \tL^*_{[2,0]} & \tL^*_{[2,1]} & \tL^*_{[2,2]} & \hdots \\
            \vdots & \vdots & \vdots & \ddots
        \end{pmatrix} \text{ and }
        \bomega^M = \begin{pmatrix}
            \bomega^M_0 \\ \bomega^M_1 \\ \bomega^M_2 \\ \vdots \\ \bomega^M_{T}
        \end{pmatrix}
    \end{align*}
    where $(M)_t$ denotes the $t^{th}$ block of matrix $M$ and the covariance matrix of $\bomega^M$ is {$\EE[\bomega^M (\bomega^M)^\top] = \diag((\Sigma/M + \Sigma^0)_{0 \leq t \leq T})$}. 
    Similarly we characterize the deviation process $x^{j*,M}_t - \bx^{M*}_t$, using \eqref{eq:epsN_x_M} and \eqref{eq:epsN_bx_M}
    \begin{align*}
        x^{j*,M}_{t+1} - \bx^{M*}_{t+1} = \underbrace{(A_t - B_t K^{1*}_t + K^{2*}_t)}_{L^*_t} (x^{j*,M}_t - \bx^{M*}_t) + \underbrace{\frac{M-1}{M}\omega^j_{t} + \frac{1}{M} \sum_{k \neq j} \omega^k_{t}}_{\bomega^{j,M}_{t}}.
    \end{align*}
    Hence 
    \begin{align*}
        x^{j*,M}_t - \bx^{M*}_t & = \sum_{s=0}^t L^*_{[t-1,s]} \bomega^{j,M}_s \\
        & = \big( \Psi^* \bomega^{j,M} \big)_t, \text{ where } \bomega^{j,M} = \frac{M-1}{M} \begin{pmatrix}
            \omega^j_0 \\ \vdots \\ \omega^j_{T-1}
        \end{pmatrix} + \frac{1}{M} \begin{pmatrix}
            \sum_{k \neq j} \omega^k_0 \\ \vdots \\ \sum_{k \neq j} \omega^k_{T-1}
        \end{pmatrix}
    \end{align*}
    where the covariance matrix of $\bomega^{j,M}$ is {$\EE[\bomega^{j,M} (\bomega^{j,M})^\top] = \diag(((M-1)/M \times \Sigma)_{0 \leq t \leq T})$}. Similarly the infinite agent limit of this process is $x^*_t - \bx^*_t = (\Psi^* \omega)_t$ where $\omega = (\omega^\top_0, \ldots, \omega^\top_{T-1})^\top$ whose covariance is {$\EE[\omega \omega^\top] = \diag((\Sigma)_{0 \leq t \leq T})$}. Now we compute the finite agent cost in terms of the noise processes,
    \begin{align*}
        & J^\gamma_M(\mcu^{*}) = \EE \bigg[ \frac{1}{M} \sum_{j \in [M]} \sum_{t = 0}^T \lVert x^{j*,M}_t - \bx^{M*}_t \rVert^2_{Q_t} + \lVert \bx^{M*}_t \rVert^2_{\bQ_t} + \lVert u^{1,j*,M}_t - \bu^{1,M*}_t \rVert^2 + \lVert \bu^{1,M*}_t \rVert^2 \\
        & \hspace{11cm} - \gamma^2 (\lVert u^{2,j*,M}_t - \bu^{2,M*}_t \rVert^2 + \lVert \bu^{2,M*}_t \rVert^2) \bigg] \\
        & = \EE \bigg[ \frac{1}{M} \sum_{j \in [M]} \sum_{t = 0}^T \lVert x^{j*,M}_t - \bx^{M*}_t \rVert^2_{Q_t + (K^{1,*}_t)^\top K^{1,*}_t - \gamma^2 (K^{2,*}_t)^\top K^{2,*}_t} + \lVert \bx^{M*}_t \rVert^2_{\bQ_t + (L^{1,*}_t)^\top L^{1,*}_t - \gamma^2 (L^{2,*}_t)^\top L^{2,*}_t}\bigg] \\
        & = \EE \bigg[ \frac{1}{M} \sum_{j \in [M]} (\Psi^* \bomega^{j,M})^\top \big( Q + (K^*)^\top R K^* \big) \Psi^* \bomega^{j,M} \\
        & \hspace{6cm} + (\bPsi^* \bomega^{M})^\top \big( \bQ + (\bK^*)^\top \bR \bK^* \big) \bPsi^* \bomega^{M} \bigg] \\
        & = \tr \big( (\Psi^*)^\top \big( Q + (K^*)^\top R K^* \big) \Psi^* \EE \big[\bomega^{j,M} (\bomega^{j,M})^\top \big] \big) \\
        & \hspace{4.5cm} + \tr \big( (\bPsi^*)^\top \big( \bQ + (\bK^*)^\top \bR \bK^* \big) \bPsi^* \EE \big[\bomega^M (\bomega^M)^\top \big] \big)
    \end{align*}
    where $Q = \diag((Q_t)_{0 \leq T})$, $R = \diag((\diag(I,-\gamma^2 I))_{0 \leq T})$ and $K^* = \diag((K^{1,*}_t,K^{2,*}_t)_{0 \leq T})$ with $K^{1,*}_T = 0$ and $K^{2,*}_T = 0$. Similarly $\bQ = \diag((\bQ_t)_{0 \leq T})$, $\bR = \diag((\diag(I,-\gamma^2 I))_{0 \leq T})$ and $\bK^* = \diag((L^{1,*}_t,L^{2,*}_t)_{0 \leq T})$ with $L^{1,*}_T = 0$ and $L^{2,*}_T = 0$. Using a similar technique we can compute the infinite agent cost:
    \begin{align*}
        J^\gamma(\mcu^{*}) & = \EE \bigg[ \sum_{t = 0}^T \lVert x^*_t - \bx^*_t \rVert^2_{Q_t} + \lVert \bx^*_t \rVert^2_{\bQ_t} + \lVert u^*_t - \bu^*_t \rVert^2 + \lVert \bu^*_t \rVert^2  \bigg] \\
        & =\tr \big( (\Psi^*)^\top \big( Q + (K^*)^\top R K^* \big) \Psi^* \EE \big[\omega \omega^\top \big] \big) \\
        & \hspace{4.5cm} + \tr \big( (\bPsi^*)^\top \big( \bQ + (\bK^*)^\top \bR \bK^* \big) \bPsi^* \EE \big[\bomega^0 (\bomega^0)^\top \big] \big).
    \end{align*}
    Now evaluating the difference between the finite and infinite population costs:
    \begin{align}
        & \lvert J^\gamma_M(\mcu^{*}) - J^\gamma(\mcu^{*}) \rvert \nonumber \\
        & =  \big|\tr \big( (\Psi^*)^\top \big( Q + (K^*)^\top R K^* \big) \Psi^* \big( \EE \big[\bomega^{j,M} (\bomega^{j,M})^\top - \EE \big[\omega \omega^\top \big]  \big) \big] \big) \nonumber \\
        & \hspace{3cm} + \tr \big( (\bPsi^*)^\top \big( \bQ + (\bK^*)^\top \bR \bK^* \big) \bPsi^* \big( \EE \big[\bomega^M (\bomega^M)^\top \big] - \EE \big[\bomega^0 (\bomega^0)^\top \big] \big) \big) \big| \nonumber \\
        & \leq \underbrace{\Big(\lVert (\Psi^*)^\top \big( Q + (K^*)^\top R K^* \big) \Psi^* \rVert_F + \lVert (\bPsi^*)^\top \big( \bQ + (\bK^*)^\top \bR \bK^* \big) \bPsi^* \rVert_F \Big)}_{C^*_1} \nonumber \\
        & \hspace{10cm} \tr \big( \diag((\Sigma/M)_{0 \leq t \leq T}) \big) \nonumber \\
        & \leq C^*_1 \frac{\sigma T}{M} \label{eq:eps_Nash_1}
    \end{align}
    where $\sigma = \lVert \Sigma \rVert_F$. Now let us consider the same dynamics but under-Nash controls. The difference between finite and infinite population costs under these controls are 
    \begin{align}
        \lvert J^\gamma_M(\mcu) - J^\gamma(\mcu) \rvert \leq \underbrace{\Big(\lVert (\Psi)^\top \big( Q + (K)^\top R K \big) \Psi \rVert_F + \lVert (\bPsi)^\top \big( \bQ + (\bK)^\top \bR \bK \big) \bPsi \rVert_F \Big)}_{C_1} \frac{\sigma T}{M}
    \end{align}
    Using this bound along we can deduce that if,
    \begin{align}
        \inf_{\mcu^1} \sup_{\mcu^2} J^\gamma (\mcu^1,\mcu^2) + C_1 \frac{\sigma T}{M} - \gamma^2 \EE \sum_{t=0}^{T-1} \lVert \omega_t \rVert^2 + \lVert \bomega_t \rVert^2  \leq 0 , \label{eq:eps_Nash_2}
    \end{align}
    then the robustness condition \eqref{eq:N_agent_inf_sup} will be satisfied. Using results form Theorem \ref{thm:NE_2P_MFTG} we can rewrite \eqref{eq:eps_Nash_2} into the following condition
    \begin{align*} 
        \sum_{t=1}^{T} \tr((M^\gamma_t - \gamma^2 I) \Sigma + (\bM^\gamma_t - \gamma^2 I )\bSigma) +\tr( M^\gamma_0 \Sigma^0) + \tr( \bM^\gamma_0 \bSigma^0) \leq -\frac{CT}{N}
    \end{align*}
    which concludes the proof.
\end{proof}

\section*{Proof of Theorem \ref{thm:inner_loop_conv}}
\begin{proof}
We start by analyzing the problem associated with the process $y^i$ and the controller $\bK_t = [(K^1_t)^\top,(K^2_t)^\top]^\top$. Using the Lyapunov equation the cost for the given set of future controllers $\big( (\tK_{t+1},\tL_{t+1}), \ldots, (\tK_T,\tL_T) \big)$ can be defined in terms of symmetric matrix $\tM_{t+1}$ and covariance matrix $\tSigma^y_t$,
\begin{align*}
	\min_{K^1_t} \max_{K^2_t} \tJ^{i,\gamma}_{y,t}(\bK_t) = \EE_y\big[ y^\top (Q_t + \bK^\top_t \bR \bK_t) y + y^\top (A-\bB_t \bK_t)^\top \tM_{t+1} (A-\bB_t \bK_t) y \big] + \tSigma^y_t,
\end{align*}
where $\bB_t = [B_t, I]$ and $\bR = \diag(I, -\gamma^2 I)$. The matrices $\tM_{t+1}$ and $\tSigma^y_t$ are fixed and can be computed offline using $\big( (\tK_{t+1},\tL_{t+1})$ $ , \ldots,(\tK_T,\tL_T) \big)$ and $\EE[\tomega^i_s (\tomega^i_s)^\top]$.
\begin{align*}
	\tM_s & = Q_t + (\tK^1_t)^\top \tK^1_t - \gamma^2 (\tK^2_t)^\top \tK^2_t + \gamma(A_t- B_t \tK^1_t + \tK^2_t)^T \tM_{s+1} (A_t- B_t \tK^1_t + \tK^2_t), \\
	\tSigma^y_s & =  \tr(\tM_{s+1} \EE[\tomega^i_s (\tomega^i_s)^\top]) + \tSigma^y_{s+1}, 
\end{align*}
for all $s \geq t+1$.
For a given $t \in \{T-1,\ldots,1,0\}$ the \emph{exact} policy gradient of cost $\tJ^{i,\gamma}_{y,t}$ with respect to $K^1_t$ and $K^2_t$ is 
\begin{align}
	\frac{\delta \tJ^{i,\gamma}_{y,t}}{\delta K^1_t} &  = 2 \big((I +  B^\top_t \tM_{t+1} B_t)K^1_t -  B^\top_t \tM_{t+1}(A_t-K^2_t)  \big) \Sigma_y, \label{eq:cost_grads}\\
	- \frac{\delta \tJ^{i,\gamma}_{y,t}}{\delta K^2_t} & = -2 \big( (-\gamma^2 I  +   \tM_{t+1} ) K^2_t +   \tM_{t+1} (A_t- B_t K^1_t) \big) \Sigma_z \nonumber
\end{align}
First we prove smoothness and convex-concave property of the function $\tJ^{i,\gamma}_{y,t}$. Due to the projection operation $\proj_D$ the norms of $K^1_t, K^2_t$ are bounded by scalar $D$. Using Definition 2.1 in \citep{fallah2020optimal} and \eqref{eq:cost_grads}, the function $ \tJ^{i,\gamma}_{y,t}$ is smooth with smoothness parameter
\begin{align*}
    L =  \big( \lVert I +  B^\top_t \tM_{t+1} B_t \rVert + \lVert \tM_{t+1} -\gamma^2 I \rVert \big) ( \lVert \Sigma_y \rVert + \lVert \Sigma_z \rVert)
\end{align*}
The function $\tJ^{i,\gamma}_{y,t}$ is also convex-concave since, using \eqref{eq:cost_grads} and convexity-concavity parameters are obtained using direct calculation,
\begin{align}
    \mu_x = \sigma((I + B^\top_t \tM_{t+1} B_t)\Sigma_y) > 0 \text{ and } \mu_y = \sigma((\tM_{t+1} - \gamma^2 I) \Sigma_z) > 0
\end{align}
where the second inequality is ensured when $\tM_{t+1} - \gamma^2 I > 0$. Moreover satisfying Assumption 2.3 in \citep{fallah2020optimal} requires component-wise smoothness which can be satisfied with $L_x = \lVert I +  B^\top_t \tM_{t+1} B_t \rVert \lVert \Sigma_y \rVert$ and $L_y = \lVert \tM_{t+1} -\gamma^2 I \rVert \lVert \Sigma_z \rVert$. Having satisfied Assumption 2.3 in \citep{fallah2020optimal} now we generalize the proof of convergence for gradient descent-ascent to \emph{biased} stochastic gradients as is the case with $ \tnabla_1 \tJ^{i,\gamma}_t$ and $\tnabla_2 \tJ^{i,\gamma}_t$.

\noindent Let us first introduce a couple of notations, $\nabla_{K_{i,t}} \tJ^{i,\gamma}(K_t,L_t)$ is exact gradient of cost $\tJ^{i,\gamma}$ w.r.t. controller $K_{i,t}$,
\begin{align*}
	\nabla_{K_{i,t}} \tJ^{i,\gamma}(K_t,L_t) = \frac{\delta \tJ^{i,\gamma}(K_t,L_t)}{ \delta K_{i,t} }
\end{align*}
$\tnabla_{K_{i,t}} \tJ^{i,\gamma}(K_t,L_t)$ is stochastic gradient of cost $\tJ^{i,\gamma}$ w.r.t. controller $K_{i,t}$,
\begin{align*}
	& \tnabla_{K_{i,t}}  \tJ^{i,\gamma}_t(K_t,L_t) = \frac{n}{Mr^2} \sum_{j=1}^M  \tJ^{i,\gamma}((K^{j,1}_t,K^2_t),L_t) e_j, \hspace{0.2cm} K^{j,1}_t = K^1_t + e_j, \hspace{0.2cm} e_j \sim \mathbb{S}^{n-1}(r),
\end{align*}
$\nabla^r_{K_{i,t}} \tJ^{i,\gamma}(K_t,L_t)$  is the smoothed gradient of cost $\tJ^{i,\gamma}$ w.r.t. controller $K_{i,t}$,
\begin{align*}
	& \text{ s.t. } \nabla^r_{K_{i,t}} \tJ^{i,\gamma}(K_t,L_t) = \frac{m}{r} \EE_e \big[ \tJ^{i,\gamma}((K_{i,t} + e, K_{-i,t}),L_t) e \big], \text{ where } e \sim \Sb^{m-1}(r)
\end{align*}
Now we introduce some results from literature. The following result proves that the stochastic gradient is an unbiased estimator of the smoothed gradient and the bias between the smoothed gradient and the stochastic gradient is bounded by a linear function of smoothing radius $r$.
\begin{lemma}[\citep{fazel2018global}]
	\begin{align*}
		\EE[\tnabla_{K_{i,t}} \tJ^{i,\gamma}(K_t,L_t)] & = \nabla^r_{K_{i,t}} \tJ^{i,\gamma}(K_t,L_t), \\
		\lVert \nabla^r_{K_{i,t}} \tJ^{i,\gamma}(K_t,L_t) - \nabla_{K_{i,t}} \tJ^{i,\gamma}(K_t,L_t) \rVert_2 & \leq \phi_i r, \text{ where } \phi_i = \lVert I +  B^\top_i M^* B_i \rVert_2
	\end{align*}
\end{lemma}
Next we present the result that difference between the \emph{mini-batched} stochastic gradient and the smoothed gradient can be bounded with high probability.
\begin{lemma}[\citep{malik2019derivative}] \label{lem:malik_hi_conf_bound}
	\begin{align*}
	\lVert \tnabla_{K_{i,t}} \tJ^{i,\gamma}(K_t,L_t) - \nabla^r_{K_{i,t}} \tJ^{i,\gamma}(K_t,L_t) \rVert_2 \leq \frac{m}{Mr} (\tJ^{i,\gamma}(K_t,L_t) + \lambda r) \sqrt{\log \bigg(\frac{2m}{\delta}\bigg)}
	\end{align*}
with probability $1-\delta$.
\end{lemma}
Now we compute finite sample guarantees for the Gradient Descent Ascent (GDA) update as given in Algorithm \ref{alg:RL^2_tP_MFTG}. For each $t \in [T]$, let us concatenate $K^1_t$ as $\bK_t = \begin{pmatrix} K^1_t \\ K^2_t \end{pmatrix}$. Let us also denote the optimal controller which optimizes $\tJ^{i,\gamma}_{y,t}$ as $\bK^*_t = \begin{pmatrix} K^{1*}_t \\ K^{2*}_t \end{pmatrix}$ such that $\tJ^{i,\gamma}_{y,t}(K^{1*}_t, K^2_t) \leq \tJ^{i,\gamma}_{y,t}(K^{1*}_t, K^{2*}_t) \leq \tJ^{i,\gamma}_{y,t}(K^1_t, K^{2*}_t)$. Since the timestep $t$ is fixed inside the inner loop of the algorithm we discard it, instead we use the iteration index $k \in [K]$. The update rule given in Algorithm \ref{alg:RL^2_tP_MFTG} is given by
\begin{align*}
	\bK_{k+1} = \bK_k -\eta_k \begin{pmatrix} -\tnabla_1 \tJ^{i,\gamma}_y(\bK_k) \\ \tnabla_2 \tJ^{i,\gamma}_y(\bK_k) \end{pmatrix} = \bK_k -\eta \tnabla\tJ^{i,\gamma}_y(\bK_k) = X \bK_k + Y \tnabla \tJ^{i,\gamma}_y(\bK_k)
\end{align*}
where $\tnabla_1 \tJ^{i,\gamma}_y(\bK_k) = \tnabla_{K^1_t} \tJ^{i,\gamma}_y(\bK_k), \tnabla_2 \tJ^{i,\gamma}_y(\bK_k) = \tnabla_{K^2_t} \tJ^{i,\gamma}_y(\bK_k), X = I$ and $Y = \eta I$. We the controller error as $\hK_k = \bK_k - \bK^*_k$, the evolution of this error is
\begin{align*}
	\hK_{k+1} = X \hK_k + Y \tnabla \tJ^{i,\gamma}(\bK_k)
\end{align*}
We define a Lyapunov function $V_p(\bK_k) := \hK^\top_k P \hK_k = p \lVert \hK_k \rVert_2^2$ for $P = p I$ for some $p > 0$.
\begin{align*}
	& \hspace{-1.5cm} V_p(\bK_{k+1}) - \rho^2 V_p(\bK_k) = (X \hK_k + Y \tnabla \tJ^{i,\gamma}(\bK_k))^\top P (X \hK_k + Y \tnabla \tJ^{i,\gamma}(\bK_k)) - \rho^2 \hK_k P \hK_k, \\
	\leq & (X \hK_k + Y \nabla \tJ^{i,\gamma}(\bK_k))^\top P (X \hK_k + Y \nabla \tJ^{i,\gamma}(\bK_k)) - \rho^2 \hK_k P \hK_k \\
	& + (Y (\tnabla \tJ^{i,\gamma}(\bK_k) - \nabla \tJ^{i,\gamma}(\bK_k)))^\top P X \hK_k + (X \hK_k)^\top P (Y (\tnabla \tJ^{i,\gamma}(\bK_k) - \nabla \tJ^{i,\gamma}(\bK_k))) \\
	& + (\tnabla \tJ^{i,\gamma}(\bK_k) - \nabla \tJ^{i,\gamma}(\bK_k))^\top Y^\top P Y (\tnabla \tJ^{i,\gamma}(\bK_k) - \nabla \tJ^{i,\gamma}(\bK_k)), \\
	\leq & \begin{pmatrix} \hK_k \\ \nabla \tJ^{i,\gamma} (\bK_k) \end{pmatrix}^\top \begin{bmatrix} X^T P X - \rho^2 P & X^\top P Y \\ Y^\top P X & Y^\top P Y \end{bmatrix} \begin{pmatrix} \hK_k \\ \nabla \tJ^{i,\gamma} (\bK_k) \end{pmatrix} + \eta^2 p \lVert \tnabla\tJ^{i,\gamma}(\bK_k) - \nabla \tJ^{i,\gamma} \rVert^2_2 \\
	& + \begin{pmatrix} \hK_k \\ \tnabla\tJ^{i,\gamma}(\bK_k) - \nabla \tJ^{i,\gamma} (\bK_k) \end{pmatrix}^\top \begin{bmatrix} 0 & X^\top P Y \\ Y^\top P X & 0 \end{bmatrix} \begin{pmatrix} \hK_k \\ \tnabla\tJ^{i,\gamma}(\bK_k) - \nabla \tJ^{i,\gamma} (\bK_k) \end{pmatrix}.
\end{align*}
Let us enforce $\rho^2 = 1 - m \eta$, then we know due to \citep{fallah2020optimal} that 
$$\begin{pmatrix} \hK_k \\ \nabla \tJ^{i,\gamma} (\bK_k) \end{pmatrix}^\top \begin{bmatrix} X^T P X - \rho^2 P & X^\top P Y \\ Y^\top P X & Y^\top P Y \end{bmatrix} \begin{pmatrix} \hK_k \\ \nabla \tJ^{i,\gamma} (\bK_k) \end{pmatrix} \leq 0$$ 
\begin{align*}
	\hK^\top_{k+1} \hK_{k+1} - (1-m\eta) \hK^\top_k \hK_k & \leq \eta \lVert \tnabla\tJ^{i,\gamma}(\bK_k) - \nabla \tJ^{i,\gamma} (\bK_k) \rVert_2 \lVert \hK_k \rVert_2 + \eta^2 \lVert \tnabla\tJ^{i,\gamma}(\bK_k) - \nabla \tJ^{i,\gamma} (\bK_k) \rVert_2^2, \\
	& = \sqrt{m \eta} \lVert \hK_k \rVert_2  \frac{\sqrt{\eta}\lVert \tnabla\tJ^{i,\gamma}(\bK_k) - \nabla \tJ^{i,\gamma} (\bK_k) \rVert_2 }{\sqrt{m}} + \eta^2 \lVert \tnabla\tJ^{i,\gamma}(\bK_k) - \nabla \tJ^{i,\gamma} (\bK_k) \rVert_2^2
\end{align*}
Using the fact $2ab \leq a^2 + b^2$ and for $\eta < 1$ we get
\begin{align*}
	\lVert \hK_{k+1} \rVert^2_2 & \leq (1-m\eta) \lVert \hK_k \rVert^2_2 +  \frac{m \eta}{2} \lVert \hK_k \rVert_2 + \eta \big(\frac{1}{m} + 1 \big) \lVert \tnabla\tJ^{i,\gamma}(\bK_k) - \nabla \tJ^{i,\gamma} (\bK_k) \rVert^2_2, \\
	& \leq \bigg(1-\frac{m\eta}{2} \bigg) \lVert \hK_k \rVert^2_2 + \eta \bigg(\frac{1}{m} + 1 \bigg) \lVert \tnabla\tJ^{i,\gamma}(\bK_k) - \nabla \tJ^{i,\gamma} (\bK_k) \rVert^2_2,
\end{align*}
Hence
\begin{align*}
	\lVert \hK_{k} \rVert^2_2 & \leq \bigg(1-\frac{m\eta}{2} \bigg)^k \lVert \hK_0 \rVert^2_2 + \eta \bigg(\frac{1}{m} + 1 \bigg) \sum_{j=0}^k \bigg( 1- \frac{m \eta}{2}\bigg)^j \lVert \tnabla\tJ^{i,\gamma}(\bK_j) - \nabla \tJ^{i,\gamma} (\bK_j) \rVert^2_2, \\
	& \leq \bigg(1-\frac{m\eta}{2} \bigg)^k \lVert \hK_0 \rVert^2_2 + \eta \bigg(\frac{1}{m} + 1 \bigg) \sum_{j=0}^\infty \bigg( 1- \frac{m \eta}{2}\bigg)^j \lVert \tnabla\tJ^{i,\gamma}(\bK_j) - \nabla \tJ^{i,\gamma} (\bK_j) \rVert^2_2, \\
	& \leq \bigg(1-\frac{m\eta}{2} \bigg)^k \lVert \hK_0 \rVert^2_2 + \frac{2}{m} \bigg(\frac{1}{m} + 1 \bigg) \bigg(\frac{m}{Mr} (\tJ^{i,\gamma}(K_t,L_t) + \lambda r) \sqrt{\log \bigg(\frac{2m}{\delta}\bigg)} + \phi_i r \bigg),
\end{align*}
If $k = \Os(\log(1/\epsilon)), M = \Os(1/\epsilon)$ and $r = \Omega(\epsilon)$ then $\lVert \hK_{k} \rVert^2_2 \leq \epsilon$.
\end{proof}

\section*{Proof of Theorem \ref{thm:main_res}}

\begin{proof}
    In this proof we show that $\max_{j \in \{1,2\}} \lVert K^j_t - K^{j*}_t \rVert = \Os(\epsilon)$ for $t \in \{T-1,\ldots,0\}$ and the result $\max_{j \in \{1,2\}} \lVert L^j_t - L^{j*}_t \rVert = \Os(\epsilon)$ can be obtained in a similar manner.  Throughout the proof we refer to the output of the inner loop of Algorithm \ref{alg:RL^2_tP_MFTG} as the set of \emph{output controllers} $(K^i_t)_{i \in [2],t \in \{0,\ldots,T-1\}}$. In the proof we use two other sets of controllers as well. The first set $(K^{i*}_t)_{i \in [2],t \in \{0,\ldots,T-1\}}$ which denotes the NE as characterized in Theorem \ref{thm:NE_2P_MFTG}. The second set is called the \emph{local}-NE (as in proof of Theorem \ref{thm:inner_loop_conv}) and is denoted by $(\tK^{i*}_t)_{i \in [2],t \in \{0,\ldots,T-1\}}$. The proof quantifies the error between the output controllers $(K^i_t)_{i \in [2]}$ and the corresponding NE controllers $(K^{i*}_t)_{i \in [2]}$ by utilizing the intermediate local-NE controllers $(\tK^{i*}_t)_{i \in [2]}$ for each time $t \in \{T-1,\ldots,0\}$. For each $t$ the error is shown to depend on error in future controllers $(K^i_s)_{s \geq t, i \in [2]}$ and the approximation error $\Delta_t$ introduced by the gradient descent-ascent update. If $\Delta_t = \Os(\epsilon)$, then the error between the output and NE controllers is shown to be $\Os(\epsilon)$.
    
    Let us start by denoting the NE value function matrices for agent $i \in [2]$ at time $t \in \{0,1,\ldots,T-1\}$, under the NE control matrices $(K^{i*}_s)_{i \in [2], s \in \{t+1,\ldots,T-1\}}$ by $M^{i*}_t$. Using results in literature \cite{bacsar1998dynamic} the local-NE control matrices can be characterized as:
    \begin{align*}
        \tK^{1*}_t = -B^\top_t M_{t+1} \Lambda^{-1}_t A_t, \hspace{0.2cm} \tK^{2*}_t = M_{t+1} \Lambda^{-1}_t A_t
    \end{align*}
    where
    \begin{align*}
       \tM_t = Q_t + A^\top_t M_{t+1} \Lambda^{-1}_t A_t, \hspace{0.2cm} \Lambda_t = I + (B_t B^\top_t - \gamma^{-2} I ) M_{t+1}
    \end{align*}
    Similarly the NE control matrices can be characterized as:
    \begin{align*}
        K^{1*}_t = -B^\top_t M^*_{t+1} {\Lambda^*_t}^{-1} A_t, \hspace{0.2cm} K^{2*}_t = M^*_{t+1} {\Lambda^*_t}^{-1} A_t
    \end{align*}
    where
    \begin{align*}
       M^*_t = Q_t + A^\top_t M^*_{t+1} {\Lambda^*_t}^{-1} A_t, \hspace{0.2cm} \Lambda^*_t = I + (B_t B^\top_t - \gamma^{-2} I ) M^*_{t+1}
    \end{align*}
    The matrix $\Lambda^*_t$ is invertible \cite{bacsar1998dynamic} and the matrix $\Lambda_t$ will also be invertible if $M_{t+1}$ is close enough to $M^*_{t+1}$. Now we characterize the difference between the local NE and NE controllers.
    \begin{align}
        K^{1*}_t - \tK^{1*}_t & = B^\top_t(M^*_{t+1} {\Lambda^*_t}^{-1} - M_{t+1} \Lambda^{-1}_t ) A_t \nonumber \\
        & = B^\top_t(M^*_{t+1} {\Lambda^*_t}^{-1} - M_{t+1} {\Lambda^*_t}^{-1} + M_{t+1} {\Lambda^*_t}^{-1} - M_{t+1} \Lambda^{-1}_t ) A_t \nonumber \\
        & = B^\top_t \big((M^*_{t+1}- M_{t+1}) {\Lambda^*_t}^{-1} + M_{t+1} ({\Lambda^*_t}^{-1} - \Lambda^{-1}_t) \big) A_t \label{eq:K^*-tK}
    \end{align}
    To characterize the difference given above we must evaluate ${\Lambda^*_t}^{-1} - \Lambda^{-1}_t$. Using matrix manipulations we can write down
    \begin{align}
        \Lambda^{-1}_t - {\Lambda^*_t}^{-1} & = \Lambda^{-1}_t \big(\Lambda^*_t - \Lambda^{-1}_t \big){\Lambda^*_t}^{-1} \nonumber \\
        & = \Lambda^{-1}_t (B_t B^\top_t - \gamma^{-2} I ) (M^*_{t+1} - M_{t+1}) {\Lambda^*_t}^{-1} \label{eq:Lam-Lam*}
    \end{align}
    Using \eqref{eq:K^*-tK} and \eqref{eq:Lam-Lam*} we can bound the difference between the local-NE and NE controllers
    \begin{align}
        \lVert K^{1*}_t - \tK^{1*}_t \rVert & = \lVert B^\top_t \big((M^*_{t+1}- M_{t+1}) {\Lambda^*_t}^{-1} + M_{t+1} ({\Lambda^*_t}^{-1} - \Lambda^{-1}_t) \big) A_t  \rVert \nonumber \\
        & \leq c_A c_B \big( 1 + c_M \lVert\Lambda^{-1}_t \rVert \big) c_\Lambda \lVert M^*_{t+1} - M_{t+1} \rVert \label{eq:K^*-tK_norm}
    \end{align}
    assuming $\lVert M^*_{t+1} - M_{t+1} \rVert \leq 1$ where $c_A := \max_t \lVert A_t \rVert, c_B := \max_t \lVert B_t \rVert, c_M := \max_t \lVert M^*_t \rVert$ and $c_\Lambda := \max_t \lVert\Lambda^{-1}_t \rVert$. Now we bound $\lVert\Lambda^{-1}_t \rVert$ by first defining $\hLambda_t := \Lambda_t - \Lambda^*_t = (B_t B^\top_t - \gamma^{-2} I ) (M_{t+1} - M^*_{t+1})$
    \begin{align}
        \lVert\Lambda^{-1}_t \rVert & = \lVert (I + {\Lambda^*_t}^{-1} \hLambda_t )^{-1} {\Lambda^*_t}^{-1} \rVert \nonumber \\
        & \leq c_\Lambda \sum_{k=0}^\infty \big(c_\Lambda \lVert \hLambda_t \rVert \big)^k \leq c_\Lambda \sum_{k=0}^\infty \big( (c^2_B + \gamma^{-2}) c_\Lambda \lVert M_{t+1} - M^*_{t+1} \rVert \big)^k \leq 2 c_\Lambda \label{eq:Lambda_t}
    \end{align}
    where the last inequality is possible due to $\lVert M_{t+1} - M^*_{t+1} \rVert \leq \frac{1}{2 c_\Lambda (c^2_B + \gamma^{-2}) }$. Using \eqref{eq:K^*-tK_norm} and \eqref{eq:Lambda_t} we arrive at
    \begin{align*}
        \lVert K^{1*}_t - \tK^{1*}_t \rVert \leq c_A c_B \big( 1 + 2 c_M c_\Lambda) c_\Lambda \lVert M^*_{t+1} - M_{t+1} \rVert
    \end{align*}
    Similarly the difference between $K^{1*}_t - \tK^{1*}_t$ can be bounded as
    \begin{align*}
        \lVert K^{2*}_t - \tK^{2*}_t \rVert \leq c_A \big( 1 + 2 c_M c_\Lambda) c_\Lambda \lVert M^*_{t+1} - M_{t+1} \rVert
    \end{align*}
    Hence we can equivalently write
    \begin{align}
        \max_{i \in \{1,2\}}\lVert K^{i*}_t - \tK^{i*}_t \rVert \leq \underbrace{c_A \max (c_B,1) \big( 1 + 2 c_M c_\Lambda) c_\Lambda}_{\bc_1} \lVert M^*_{t+1} - M_{t+1} \rVert \label{eq:tKi_K*}
    \end{align}

    Having bounded the difference between local-NE and the NE controllers at time $t$, now we turn towards bounding the difference between the value matrices $M^*_t$ and $M_t$. First we define a change of notation to keep the analysis concise. Let
    \begin{align*}
        B^1_t = B_t, \hspace{0.2cm} B^2_t = I, \hspace{0.2cm} Q^1_t = Q_t, \hspace{0.2cm} Q^2_t = -Q_t, \hspace{0.2cm} R^1_t = I, \hspace{0.2cm} R^2_t = -\gamma^2 I, \hspace{0.2cm} M^1_t = M_t, \hspace{0.2cm} M^2_t = -M_t
    \end{align*}
    where the sequence $(M_t)_{\forall t}$ will be defined in the following analysis. 
    The NE value function matrices are be defined recursively using the Lyapunov equation and NE controllers as 
    \begin{align}
        M^{i*}_t & = Q^i_t + (A^{i*}_t)^\top M^{i*}_{t+1} A^{i*}_t - (A^{i*}_t)^\top M^{i*}_{t+1} B^i_t (R^i_t + (B^i_t)^\top M^{i*}_{t+1} B^i_t)^{-1} (B^i_t)^\top M^{i*}_{t+1} A^{i*}_t \nonumber \\
        & = Q^i_t + (A^{i*}_t)^\top M^{i*}_{t+1} ( A^{i*}_t + B^i_t K^{i*}_t), \nonumber \\
        M^{i*}_T & = Q^i_T \label{eq:Pi*t}
    \end{align}
    where $A^{i*}_t := A_t + \sum_{j \neq i} B^j_t K^{j*}_t$. The sufficient condition for existence and uniqueness of the set of matrices $K^{i*}_t$ and $M^{i*}_t$ is shown in Theorem \ref{thm:NE_2P_MFTG}. Let us now define the perturbed values matrices $\tM^i_t$ (resulting from the control matrices $(K^i_s)_{i \in [2], s \in \{t+1,\ldots,T-1\}}$). 
    Using these matrices we define the value function matrices $(\tM^{i}_t)_{i \in [2]}$ using the Lyapunov equations as follows
    \begin{align}
        \tM^i_t & =  Q^i_t + (\tK^{i*}_t)^\top R^i_t \tK^{i*}_t + \bigg(A_t + \sum_{j=1}^N B^j_t \tK^{i*}_t \bigg)^\top M^i_{t+1} \bigg(A_t + \sum_{j=1}^2 B^{i}_t \tK^{j*}_t \bigg) \nonumber \\
        & = Q^i_t + (\tA^i_t)^\top M^i_{t+1} \tA^i_t - (\tA^i_t)^\top M^i_{t+1} B^i_t (R^i_t + (B^i_t)^\top M^i_{t+1} B^i_t)^{-1} (B^i_t)^\top M^i_{t+1} \tA^i_t \nonumber \\
        & = Q^i_t + (\tA^i_t)^\top M^i_{t+1} ( \tA^i_t + B^i_t \tK^{i*}_t) \nonumber \\
        \tM^i_T & = Q^i_T \label{eq:tPit}
    \end{align}
    where $\tK^{j*} = (\tK^{j*}_t, K^j_{t+1},\ldots,K^j_{T-1})$ and $\tA^i_t := A_t + \sum_{j \neq i} B^j_t \tK^{j*}_t$. Finally we define the perturbed value function matrices $M^i_t$ which result from the perturbed matrices $(K^i_s)_{i \in [2], s \in \{t+1,\ldots,T-1\}}$ obtained using the gradient descent-ascent in Algorithm \ref{alg:RL^2_tP_MFTG}:
    \begin{align}
        M^i_t = Q^i_t + (K^i_t)^\top R^i_t K^i_t + \bigg(A_t + \sum_{j=1}^2 B^j_t K^j_t \bigg)^\top M^i_{t+1} \bigg(A_t + \sum_{j=1}^2 B^j_t K^j_t \bigg) \label{eq:Pit}
    \end{align}
    Throughout this proof we assume that the output of the inner loop in Algorithm \ref{alg:RL^2_tP_MFTG}, also called the \emph{output matrices} $K^i_t$,
    are $\Delta_t$ away from the target matrices $\tK^{i*}_t$, such that $\lVert K^i_t - \tK^{i*}_t \rVert \leq \Delta_t$. We know that,
    \begin{align*}
        \lVert K^i_t - K^{i*}_t \rVert \leq \lVert K^i_t - \tK^{i*}_t \rVert + \lVert \tK^{i*}_t - K^{i*}_t \rVert \leq \Delta_t + \lVert \tK^{i*}_t - K^{i*}_t \rVert.
    \end{align*}

    Now we characterize the difference $M^i_t - M^{i*}_t = M^i_t - \tM^i_t + \tM^i_t - M^{i*}_t$. First we can characterize $\tM^i_t - M^{i*}_t$ using \eqref{eq:Pi*t} and \eqref{eq:tPit}:
    \begin{align}
        & \tM^i_t - M^{i*}_t = (\tA^i_t)^\top M^i_{t+1} ( \tA^i_t + B^i_t \tK^{i*}_t) - (A^{i*}_t)^\top M^{i*}_{t+1} ( A^{i*}_t + B^i_t K^{i*}_t) \nonumber \\
        & = (\tA^i_t - A^{i*}_t)^\top M^i_{t+1} ( \tA^i_t + B^i_t \tK^{i*}_t) + (A^{i*}_t)^\top (M^i_{t+1} - M^{i*}_{t+1}) (\tA^i_t + B^i_t \tK^{i*}_t) \nonumber \\
        & \hspace{6cm} + (A^{i*}_t)^\top M^{i*}_{t+1} ( \tA^i_t + B^i_t \tK^{i*}_t - A^{i*}_t - B^i_t K^{i*}_t) \nonumber \\
        & = \bigg(\sum_{j \neq i} B^j_t (\tK^{j*}_t - K^{j*}_t) \bigg)^\top M^i_{t+1} ( \tA^i_t + B^i_t \tK^{i*}_t) + (A^{i*}_t)^\top (M^i_{t+1} - M^{i*}_{t+1}) (\tA^i_t + B^i_t \tK^{i*}_t) \nonumber \\
        & \hspace{6cm} + (A^{i*}_t)^\top M^{i*}_{t+1} \sum_{j =1}^2 B^j_t (\tK^{j*}_t - K^{j*}_t)
    \end{align}
    Using this characterization we bound $\lVert \tM^i_t - M^{i*}_t \rVert$ using the AM-GM inequality
    \begin{align}
        & \lVert \tM^i_t - M^{i*}_t \rVert \nonumber \\
        & \leq 2 \lVert A^*_t \rVert \lVert M^{i*}_{t+1} \rVert c_B \sum_{j=1}^2 \lVert \tK^{j*}_t - K^{j*}_t \rVert \nonumber \\
        & \hspace{0.5cm} + \big(\lVert A^*_t \rVert/2 + \lVert M^{i*}_{t+1} \rVert c_B + \lVert A^{i*}_t \rVert/2 \big) c_B \bigg(\sum_{j=1}^2 \lVert \tK^{j*}_t - K^{j*}_t \rVert \bigg)^2 + \frac{c^4_B}{2} \bigg( \sum_{j=1}^2 \lVert \tK^{j*}_t - K^{j*}_t \rVert \bigg)^4  \nonumber \\
        & \hspace{0.5cm} + \big(\lVert A^*_t \rVert/2 + 1/2 + \lVert A^{i*}_t \rVert/2 \big) \lVert M^i_{t+1} - M^{i*}_{t+1} \rVert^2 + \lVert A^{i*}_t \rVert \lVert A^*_t \rVert \lVert M^i_{t+1} - M^{i*}_{t+1} \rVert \nonumber \\
        & \leq \underbrace{\big( 2 c^*_A c^*_M c_B + (c^*_A /2 + c^*_M c_B + c^{i*}_A/2 ) c_B + c^4_B/2 \big)}_{\bc_2} \sum_{j=1}^2 \lVert \tK^{j*}_t - K^{j*}_t \rVert \nonumber \\
        & \hspace{7cm} + \underbrace{c^*_A/2 + 1/2 + c^{i*}_A/2 + c^{i*}_A c^*_A}_{\bc_3} \lVert M^i_{t+1} - M^{i*}_{t+1} \rVert \nonumber \\
        & \leq \bc_2 N \max_{j \in [2]} \lVert \tK^{j*}_t - K^{j*}_t \rVert + \bc_3 \lVert M^i_{t+1} - M^{i*}_{t+1} \rVert \label{eq:tPi_Pi*}
    \end{align}
    where { $A^*_t = A^{i*}_t + B^i_t K^{i*}_t, c^*_A = \lVert A^*_t \rVert, c^{i*}_A := \max_{t \in \{0,\ldots,T-1\}} \lVert A^{i*}_t \rVert, c^*_M := \max_{i \in [2], t \in \{0,\ldots,T-1\}} \lVert M^{i*}_{t+1} \rVert$}, and the last inequality is possible due to the fact that { $\lVert M^i_{t+1} - M^{i*}_{t+1} \rVert, \lVert \tK^{j*}_t - K^{j*}_t \rVert \leq 1/N$}. Similarly $M^i_t - \tM^i_t$ can be decomposed using \eqref{eq:tPit} and \eqref{eq:Pit}:
    \begin{align*}
        M^i_t - \tM^i_t & = (K^i_t)^\top R^i_t K^i_t + \bigg(A_t + \sum_{j=1}^2 B^j_t K^j_t \bigg)^\top M^i_{t+1} \bigg(A_t + \sum_{j=1}^2 B^j_t K^j_t \bigg) \\
        & \hspace{2cm} - \bigg[ (\tK^{i*}_t)^\top R^i_t \tK^{i*}_t + \bigg(A_t + \sum_{j=1}^2 B^j_t \tK^{i*}_t \bigg)^\top M^i_{t+1} \bigg(A_t + \sum_{j=1}^2 B^{i}_t \tK^{j*}_t \bigg) \bigg].
    \end{align*}
    We start by analyzing the quadratic form $x^\top M^i_t x$:
    \begin{align*}
        x^\top M^i_t x & = x^\top \Bigg[ Q^i_t + (K^i_t)^\top R^i_t K^i_t + \bigg(A_t + \sum_{j=1}^2 B^j_t K^j_t \bigg)^\top M^i_{t+1} \bigg(A_t + \sum_{j=1}^2 B^j_t K^j_t \bigg) \bigg] x \\
        & = x^\top \bigg[ (K^i_t)^\top \big(R^i_t + (B^i_t)^\top M^i_{t+1} B^i_t \big) K^i_t + 2 (B^i_t K^i_t)^\top M^i_{t+1} \bigg( A_t + \sum_{j \neq i} B^j_t K^j_t \bigg) + Q^i_t \\
        & \hspace{4cm} + \bigg( A_t + \sum_{j \neq i} B^j_t K^j_t \bigg)^\top M^i_{t+1} \bigg( A_t + \sum_{j \neq i} B^j_t K^j_t \bigg)\bigg] x \\
        & = x^\top \bigg[ (K^i_t)^\top \big(R^i_t + (B^i_t)^\top M^i_{t+1} B^i_t \big) K^i_t + 2 (B^i_t K^i_t)^\top M^i_{t+1} \bigg( A_t + \sum_{j \neq i} B^j_t \tK^{j*}_t \bigg) + Q^i_t \\
        & \hspace{0.4cm} + 2 (B^i_t K^i_t)^\top M^i_{t+1} \sum_{j \neq i} B^j_t (K^j_t - \tK^{j*}_t)  + \bigg( A_t + \sum_{j \neq i} B^j_t \tK^{j*}_t \bigg)^\top M^i_{t+1} \bigg( A_t + \sum_{j \neq i} B^j_t \tK^{j*}_t \bigg) \\
        & \hspace{0.4cm} + \bigg( \sum_{j \neq i} B^j_t (K^j_t - \tK^{j*}_t) \bigg)^\top M^i_{t+1} \bigg( \sum_{j \neq i} B^j_t (K^j_t - \tK^{j*}_t) \bigg) \\
        & \hspace{0.4cm} + 2 \bigg( A_t + \sum_{j \neq i} B^j_t \tK^{j*}_t \bigg) M^i_{t+1} \bigg( \sum_{j \neq i} B^j_t (K^j_t - \tK^{j*}_t) \bigg) \bigg] x
    \end{align*}
    Completing squares we get
    \begin{align}
        & x^\top M^i_t x  \label{eq:xPix} \\
        & = x^\top \Bigg[ (K^i_t - \tK^{i*}_t)^\top \big(R^i_t + (B^i_t )^\top M^i_{t+1} B^i_t \big) (K^i_t - \tK^{i*}_t)  \nonumber \\
        & \hspace{0.4cm} + \bigg( A_t + \sum_{j \neq i} B^j_t \tK^{j*}_t \bigg)^\top M^i_{t+1} \bigg( A_t + \sum_{j \neq i} B^j_t \tK^{j*}_t \bigg) + Q^i_t \nonumber \\
        & \hspace{0.4cm} - \bigg( A_t + \sum_{j \neq i} B^j_t \tK^{j*}_t \bigg)^\top M^i_{t+1} B^i_t  \big(R^i_t + (B^i_t )^\top M^i_{t+1} B^i_t \big)^{-1} (B^i_t)^\top M^i_{t+1} \bigg( A_t + \sum_{j \neq i} B^j_t \tK^{j*}_t \bigg) \nonumber \\
        & \hspace{0.4cm} + \bigg(2 \bigg(A_t + \sum_{j = 1}^2 B^j_t \tK^{j*}_t \bigg) + \sum_{j=1}^2 B^j_t (K^j_t - \tK^{j*}_t ) \bigg)^\top M^i_{t+1} \bigg( \sum_{j \neq i} B^j_t (K^j_t - \tK^{j*}_t \bigg) \bigg] x. \nonumber
    \end{align}
    Now we take a look at the quadratic $ x^\top \tM^i_t x$:
    \begin{align}
        & x^\top \tM^i_t x \nonumber \\
        & = x^\top \bigg[ Q^i_t + (\tK^{i*}_t)^\top R^i_t \tK^{i*}_t + \bigg(A_t + \sum_{j=1}^2 B^j_t \tK^{j*}_t \bigg)^\top M^i_{t+1} \bigg(A_t + \sum_{j=1}^2 B^{j}_t \tK^{j*}_t \bigg) \bigg] x \nonumber \\ 
        & = x^\top \bigg[ (\tK^{i*}_t)^\top \big( R^i_t + (B^i_t)^\top M^i_{t+1} B^i_t \big) \tK^{i*}_t + 2 (B^i_t \tK^{i*}_t)^\top M^i_{t+1} \bigg(A_t + \sum_{j \neq i} B^j_t \tK^{j*}_t \bigg) + Q^i_t \nonumber \\
        & \hspace{0.4cm} + \bigg(A_t + \sum_{j \neq i} B^j_t \tK^{j*}_t \bigg)^\top M^i_{t+1} \bigg(A_t + \sum_{j \neq i}^2 B^{j}_t \tK^{j*}_t \bigg) \bigg] x \nonumber \\
        & = x^\top \bigg[ \bigg(A_t + \sum_{j \neq i} B^j_t \tK^{j*}_t \bigg)^\top M^i_{t+1} \bigg(A_t + \sum_{j \neq i}^2 B^{j}_t \tK^{j*}_t \bigg) + Q^i_t \label{eq:xtPix} \\
        & \hspace{0.4cm}  - \bigg( A_t + \sum_{j \neq i} B^j_t \tK^{j*}_t \bigg)^\top M^i_{t+1} B^i_t  \big(R^i_t + (B^i_t )^\top M^i_{t+1} B^i_t \big)^{-1} (B^i_t)^\top M^i_{t+1} \bigg( A_t + \sum_{j \neq i} B^j_t \tK^{j*}_t \bigg) \bigg] x. \nonumber 
    \end{align}
    Using \eqref{eq:xPix} and \eqref{eq:xtPix}, we get
    \begin{align*}
        & x^\top (M^i_t - \tM^i_t) x \\
        & = x^\top \Bigg[ (K^i_t - \tK^{i*}_t)^\top \big(R^i_t + (B^i_t )^\top M^i_{t+1} B^i_t \big) (K^i_t - \tK^{i*}_t) \nonumber \\
        & \hspace{0.4cm} + \bigg(2 \bigg(A_t + \sum_{j = 1}^2 B^j_t \tK^{j*}_t \bigg) + \sum_{j=1}^2 B^j_t (K^j_t - \tK^{j*}_t ) \bigg)^\top M^i_{t+1} \bigg( \sum_{j \neq i} B^j_t (K^j_t - \tK^{j*}_t \bigg) \bigg] x \\
        & = x^\top \Bigg[ (K^i_t - \tK^{i*}_t)^\top \big(R^i_t + (B^i_t )^\top M^i_{t+1} B^i_t \big) (K^i_t - \tK^{i*}_t) \nonumber \\
        & \hspace{0.4cm} + \bigg(2 \bigg(A_t + \sum_{j = 1}^2 B^j_t K^{j*}_t \bigg) + 2 \sum_{j=1}^2 B^j_t (\tK^{j*}_t - K^{j*}_t ) + \sum_{j=1}^2 B^j_t (K^j_t - \tK^{j*}_t ) \bigg)^\top \\
        & \hspace{9cm} M^i_{t+1} \bigg( \sum_{j \neq i} B^j_t (K^j_t - \tK^{j*}_t \bigg) \bigg] x
    \end{align*}
    Using this characterization we bound $\lVert M^i_t - \tM^i_t \rVert$:
    \begin{align*}
        & \lVert M^i_t - \tM^i_t \rVert \\
        & \leq \big( \lVert R^i_t + (B^i_t)^\top M^{i*}_{t+1} B^i_t \rVert + \lVert (B^i_t)^\top \big( M^{i*}_{t+1} - M^{i*}_{t+1} \big) B^i_t \rVert \big) \lVert K^i_t - \tK^{i*}_t \rVert \\
        & \hspace{0.4cm} + (2 c^{i*}_A + 2 c_B \sum_{j=1}^2 \big( \lVert \tK^{j*}_t - K^*_t \rVert + \lVert K^j_t - \tK^{j*}_t \rVert \big) \big( \lVert M^{i*}_{t+1} \rVert + \lVert M^i_{t+1} - M^{i*}_{t+1} \rVert \big) \\
        & \hspace{9cm}c_B \sum_{j=1}^2 \lVert K^j_t - \tK^{j*}_t \rVert
    \end{align*}
    As before assuming { $\lVert M^i_{t+1} - M^{i*}_{t+1} \rVert, \lVert \tK^{j*}_t - K^{j*}_t \rVert \leq 1/N$},
    \begin{align}
        \lVert M^i_t - \tM^i_t \rVert & \leq \underbrace{\big( \bc^* + c^2_B/2 + 2 c^2_B (c^*_M + 1) + 2 c^{i*}_A \big)}_{\bc_4} \sum_{j=1}^2 \lVert K^j_t - \tK^{j*}_t \rVert  \nonumber \\
        & \hspace{1.5cm} + \underbrace{\big( c^2_B/2 + c^{i*}_A + c_B \big)}_{\bc_5} \lVert M^i_{t+1} - M^{i*}_{t+1} \rVert + \underbrace{c_B (c^*_M + 1)}_{\bc_6} \sum_{j=1}^2 \lVert \tK^{j*}_t - K^{j*}_t \rVert \nonumber \\
        & \leq \bc_4 N \Delta_t + \bc_5 \lVert M^i_{t+1} - M^{i*}_{t+1} \rVert + \bc_6 \sum_{j=1}^2 \lVert \tK^{j*}_t - K^{j*}_t \rVert \label{eq:Pi_tPi}
    \end{align}
    where { $\bc^* := \max_{i \in [2], t \in \{0,\ldots,T-1\}} \lVert R^i_t + (B^i_t)^\top M^{i*}_{t+1} B^i_t \rVert$}. Let us define  $e^K_t := \max_{j \in [2]} \lVert K^j_t - K^{j*}_t \rVert, e^P_t := \max_{j \in [2]} \lVert M^j_t - M^{j*}_t \rVert$. Using \eqref{eq:tKi_K*}, \eqref{eq:tPi_Pi*} and \eqref{eq:Pi_tPi} we get
    \begin{align*}
        e^K_t & \leq \bc_1 e^P_{t+1} + \Delta_t \\
        e^P_t & \leq (\bc_2 + \bc_6) N e^K_t + (\bc_3 + \bc_5) e^P_{t+1} + \bc_4 N \Delta \\
        & \leq \underbrace{(\bc_1 (\bc_2 + \bc_6) N + \bc_3 + \bc_5)}_{\bc_7} e^P_{t+1} + \underbrace{(\bc_2 + \bc_4 + \bc_6)N}_{\bc_8} \Delta_t
    \end{align*}
    Using this recursive definition we deduce
    \begin{align*}
        e^P_t & \leq \bc^{T-t}_7 e^P_T + \bc_8 \sum_{s=0}^{T-1} \bc^s_7  \Delta_{t+s} = \bc_8 \sum_{s=0}^{T-1} \bc^s_7  \Delta_{t+s}
    \end{align*}
    Hence if $\Delta = \Os(\epsilon)$, in particular $\Delta_t \leq \epsilon/(2 \bc_1 \bc^t_7 \bc_8 T)$ then $e^P_t \leq \epsilon/2 \bc_1$ and $e^K_t \leq \epsilon$ for $t \in \{ 0, \ldots, T-1 \}$.
\end{proof}

%% file: 2PZS-LQ-MFTG.bbl
\begin{thebibliography}{65}
\providecommand{\natexlab}[1]{#1}
\providecommand{\url}[1]{\texttt{#1}}
\expandafter\ifx\csname urlstyle\endcsname\relax
  \providecommand{\doi}[1]{doi: #1}\else
  \providecommand{\doi}{doi: \begingroup \urlstyle{rm}\Url}\fi

\bibitem[Anahtarci et~al.(2023)Anahtarci, Kariksiz, and Saldi]{anahtarci2023q}
Berkay Anahtarci, Can~Deha Kariksiz, and Naci Saldi.
\newblock Q-learning in regularized mean-field games.
\newblock \emph{Dynamic Games and Applications}, 13\penalty0 (1):\penalty0 89--117, 2023.

\bibitem[Angiuli et~al.(2022)Angiuli, Fouque, and Lauri{\`e}re]{angiuli2022unified}
Andrea Angiuli, Jean-Pierre Fouque, and Mathieu Lauri{\`e}re.
\newblock Unified reinforcement {Q}-learning for mean field game and control problems.
\newblock \emph{Mathematics of Control, Signals, and Systems}, pages 1--55, 2022.

\bibitem[Athans et~al.(1977)Athans, Castanon, Dunn, Greene, Lee, Sandell, and Willsky]{athans1977stochastic}
Michael Athans, David Castanon, K-P Dunn, C~Greene, Wing Lee, N~Sandell, and A~Willsky.
\newblock The stochastic control of the f-8c aircraft using a multiple model adaptive control (mmac) method--part i: Equilibrium flight.
\newblock \emph{IEEE Transactions on Automatic Control}, 22\penalty0 (5):\penalty0 768--780, 1977.

\bibitem[Ba{\c{s}}ar(1989)]{bacsar1989dynamic}
Tamer Ba{\c{s}}ar.
\newblock A dynamic games approach to controller design: Disturbance rejection in discrete time.
\newblock In \emph{Proceedings of the 28th IEEE Conference on Decision and Control,}, pages 407--414. IEEE, 1989.

\bibitem[Ba{\c{s}}ar and Bernhard(2008)]{bacsar2008h}
Tamer Ba{\c{s}}ar and Pierre Bernhard.
\newblock \emph{H-infinity optimal control and related minimax design problems: a dynamic game approach}.
\newblock Springer Science \& Business Media, 2008.

\bibitem[Ba{\c{s}}ar and Olsder(1998)]{bacsar1998dynamic}
Tamer Ba{\c{s}}ar and Geert~Jan Olsder.
\newblock \emph{Dynamic noncooperative game theory}.
\newblock SIAM, 1998.

\bibitem[Bensoussan et~al.(2013)Bensoussan, Frehse, Yam, et~al.]{bensoussan2013mean}
Alain Bensoussan, Jens Frehse, Phillip Yam, et~al.
\newblock \emph{Mean field games and mean field type control theory}, volume 101.
\newblock Springer, 2013.

\bibitem[Cardaliaguet and Lehalle(2018)]{cardaliaguet2018mean}
Pierre Cardaliaguet and Charles-Albert Lehalle.
\newblock Mean field game of controls and an application to trade crowding.
\newblock \emph{Mathematics and Financial Economics}, 12\penalty0 (3):\penalty0 335--363, 2018.

\bibitem[Carmona and Delarue(2018)]{carmona2018prob}
Rene Carmona and Fran{\c{c}}ois Delarue.
\newblock \emph{Probabilistic Theory of Mean Field Games with Applications I}.
\newblock Springer, Cham, 2018.

\bibitem[Carmona et~al.(2019{\natexlab{a}})Carmona, Lauri{\`e}re, and Tan]{carmona2019linear}
Ren{\'e} Carmona, Mathieu Lauri{\`e}re, and Zongjun Tan.
\newblock Linear-quadratic mean-field reinforcement learning: convergence of policy gradient methods.
\newblock \emph{arXiv preprint arXiv:1910.04295}, 2019{\natexlab{a}}.

\bibitem[Carmona et~al.(2019{\natexlab{b}})Carmona, Lauri{\`e}re, and Tan]{carmona2019model}
Ren{\'e} Carmona, Mathieu Lauri{\`e}re, and Zongjun Tan.
\newblock Model-free mean-field reinforcement learning: mean-field {MDP} and mean-field {Q}-learning.
\newblock \emph{arXiv preprint arXiv:1910.12802}, 2019{\natexlab{b}}.

\bibitem[Carmona et~al.(2020)Carmona, Hamidouche, Lauri{\`e}re, and Tan]{carmona2020policy}
Ren{\'e} Carmona, Kenza Hamidouche, Mathieu Lauri{\`e}re, and Zongjun Tan.
\newblock Policy optimization for linear-quadratic zero-sum mean-field type games.
\newblock In \emph{2020 59th IEEE Conference on Decision and Control (CDC)}, pages 1038--1043. IEEE, 2020.

\bibitem[Carmona et~al.(2021)Carmona, Hamidouche, Lauri{\`e}re, and Tan]{carmona2021linear}
Ren{\'e} Carmona, Kenza Hamidouche, Mathieu Lauri{\`e}re, and Zongjun Tan.
\newblock Linear-quadratic zero-sum mean-field type games: Optimality conditions and policy optimization.
\newblock \emph{Journal of Dynamics \& Games}, 8\penalty0 (4), 2021.

\bibitem[Choutri et~al.(2019)Choutri, Djehiche, and Tembine]{choutri2019optimal}
Salah~Eddine Choutri, Boualem Djehiche, and Hamidou Tembine.
\newblock Optimal control and zero-sum games for markov chains of mean-field type.
\newblock \emph{Mathematical Control and Related Fields}, 9\penalty0 (3):\penalty0 571--605, 2019.

\bibitem[Cosso and Pham(2019)]{cosso2019zero}
Andrea Cosso and Huy{\^e}n Pham.
\newblock Zero-sum stochastic differential games of generalized mckean--vlasov type.
\newblock \emph{Journal de Math{\'e}matiques Pures et Appliqu{\'e}es}, 129:\penalty0 180--212, 2019.

\bibitem[Cui and Koeppl(2021{\natexlab{a}})]{cui2021approximately}
Kai Cui and Heinz Koeppl.
\newblock Approximately solving mean field games via entropy-regularized deep reinforcement learning.
\newblock In \emph{International Conference on Artificial Intelligence and Statistics}, pages 1909--1917. PMLR, 2021{\natexlab{a}}.

\bibitem[Cui and Koeppl(2021{\natexlab{b}})]{cui2021learning}
Kai Cui and Heinz Koeppl.
\newblock Learning graphon mean field games and approximate {N}ash equilibria.
\newblock 2021{\natexlab{b}}.

\bibitem[Elie et~al.(2020)Elie, Perolat, Lauri{\`e}re, Geist, and Pietquin]{elie2020convergence}
Romuald Elie, Julien Perolat, Mathieu Lauri{\`e}re, Matthieu Geist, and Olivier Pietquin.
\newblock On the convergence of model free learning in mean field games.
\newblock 34\penalty0 (05):\penalty0 7143--7150, 2020.

\bibitem[Fabian et~al.(2023)Fabian, Cui, and Koeppl]{fabian2023learning}
Christian Fabian, Kai Cui, and Heinz Koeppl.
\newblock Learning sparse graphon mean field games.
\newblock In \emph{International Conference on Artificial Intelligence and Statistics}, pages 4486--4514. PMLR, 2023.

\bibitem[Fallah et~al.(2020)Fallah, Ozdaglar, and Pattathil]{fallah2020optimal}
Alireza Fallah, Asuman Ozdaglar, and Sarath Pattathil.
\newblock An optimal multistage stochastic gradient method for minimax problems.
\newblock In \emph{2020 59th IEEE Conference on Decision and Control (CDC)}, pages 3573--3579. IEEE, 2020.

\bibitem[Fazel et~al.(2018)Fazel, Ge, Kakade, and Mesbahi]{fazel2018global}
Maryam Fazel, Rong Ge, Sham~M Kakade, and Mehran Mesbahi.
\newblock Global convergence of policy gradient methods for the linear quadratic regulator.
\newblock In \emph{International Conference on Machine Learning}, pages 1467--1476, 2018.

\bibitem[Gu et~al.(2021)Gu, Guo, Wei, and Xu]{gu2021mean}
Haotian Gu, Xin Guo, Xiaoli Wei, and Renyuan Xu.
\newblock Mean-field controls with {Q}-learning for cooperative {MARL}: convergence and complexity analysis.
\newblock \emph{SIAM Journal on Mathematics of Data Science}, 3\penalty0 (4):\penalty0 1168--1196, 2021.

\bibitem[Guan et~al.(2024)Guan, Afshari, and Tsiotras]{guan2024zero}
Yue Guan, Mohammad Afshari, and Panagiotis Tsiotras.
\newblock Zero-sum games between mean-field teams: Reachability-based analysis under mean-field sharing.
\newblock In \emph{Proceedings of the AAAI Conference on Artificial Intelligence}, volume~38, pages 9731--9739, 2024.

\bibitem[Guo et~al.(2019)Guo, Hu, Xu, and Zhang]{guo2019learning}
Xin Guo, Anran Hu, Renyuan Xu, and Junzi Zhang.
\newblock Learning mean-field games.
\newblock In \emph{Advances in Neural Information Processing Systems}, 2019.

\bibitem[Harvey and Stein(1978)]{harvey1978quadratic}
Charles Harvey and Gunter Stein.
\newblock Quadratic weights for asymptotic regulator properties.
\newblock \emph{IEEE Transactions on Automatic Control}, 23\penalty0 (3):\penalty0 378--387, 1978.

\bibitem[He et~al.(2023)He, Han, Su, Han, Zou, and Miao]{he2023robust}
Sihong He, Songyang Han, Sanbao Su, Shuo Han, Shaofeng Zou, and Fei Miao.
\newblock Robust multi-agent reinforcement learning with state uncertainty.
\newblock \emph{Transactions on Machine Learning Research}, 2023.

\bibitem[Huang et~al.(2003)Huang, Caines, and Malham{\'e}]{huang2003individual}
Minyi Huang, Peter~E Caines, and Roland~P Malham{\'e}.
\newblock Individual and mass behaviour in large population stochastic wireless power control problems: Centralized and {N}ash equilibrium solutions.
\newblock In \emph{IEEE International Conference on Decision and Control}, volume~1, pages 98--103. IEEE, 2003.

\bibitem[Huang et~al.(2006)Huang, Malham{\'e}, and Caines]{huang2006large}
Minyi Huang, Roland~P Malham{\'e}, and Peter~E Caines.
\newblock Large population stochastic dynamic games: Closed-loop {M}ckean-{V}lasov systems and the {N}ash certainty equivalence principle.
\newblock \emph{Communications in Information \& Systems}, 6\penalty0 (3):\penalty0 221--252, 2006.

\bibitem[Kober et~al.(2013)Kober, Bagnell, and Peters]{kober2013reinforcement}
Jens Kober, J~Andrew Bagnell, and Jan Peters.
\newblock Reinforcement learning in robotics: A survey.
\newblock \emph{The International Journal of Robotics Research}, 32\penalty0 (11):\penalty0 1238--1274, 2013.

\bibitem[Kos and Song(2017)]{kos2017delving}
Jernej Kos and Dawn Song.
\newblock Delving into adversarial attacks on deep policies.
\newblock \emph{arXiv preprint arXiv:1705.06452}, 2017.

\bibitem[Lasry and Lions(2006)]{lasry2006jeux}
Jean-Michel Lasry and Pierre-Louis Lions.
\newblock Jeux {\`a} champ moyen. i--le cas stationnaire.
\newblock \emph{Comptes Rendus Math{\'e}matique}, 343\penalty0 (9):\penalty0 619--625, 2006.

\bibitem[Lauri{\`e}re et~al.(2022{\natexlab{a}})Lauri{\`e}re, Perrin, Girgin, Muller, Jain, Cabannes, Piliouras, P{\'e}rolat, Elie, Pietquin, et~al.]{lauriere2022scalable}
Mathieu Lauri{\`e}re, Sarah Perrin, Sertan Girgin, Paul Muller, Ayush Jain, Theophile Cabannes, Georgios Piliouras, Julien P{\'e}rolat, Romuald Elie, Olivier Pietquin, et~al.
\newblock Scalable deep reinforcement learning algorithms for mean field games.
\newblock In \emph{International Conference on Machine Learning}, pages 12078--12095. PMLR, 2022{\natexlab{a}}.

\bibitem[Lauri{\`e}re et~al.(2022{\natexlab{b}})Lauri{\`e}re, Perrin, P\'erolat, Girgin, Muller, \'Elie, Geist, and Pietquin]{lauriere2022learning}
Mathieu Lauri{\`e}re, Sarah Perrin, Julien P\'erolat, Sertan Girgin, Paul Muller, Romuald \'Elie, Matthieu Geist, and Olivier Pietquin.
\newblock Learning mean field games: A survey.
\newblock \emph{arXiv preprint arXiv:2205.12944}, 2022{\natexlab{b}}.

\bibitem[Li et~al.(2019)Li, Wu, Cui, Dong, Fang, and Russell]{li2019robust}
Shihui Li, Yi~Wu, Xinyue Cui, Honghua Dong, Fei Fang, and Stuart Russell.
\newblock Robust multi-agent reinforcement learning via minimax deep deterministic policy gradient.
\newblock In \emph{Proceedings of the AAAI conference on artificial intelligence}, volume~33, pages 4213--4220, 2019.

\bibitem[Li et~al.(2021)Li, Tang, Zhang, and Li]{li2021distributed}
Yingying Li, Yujie Tang, Runyu Zhang, and Na~Li.
\newblock Distributed reinforcement learning for decentralized linear quadratic control: A derivative-free policy optimization approach.
\newblock \emph{IEEE Transactions on Automatic Control}, 67\penalty0 (12):\penalty0 6429--6444, 2021.

\bibitem[Lowe et~al.(2017)Lowe, Wu, Tamar, Harb, Pieter~Abbeel, and Mordatch]{lowe2017multi}
Ryan Lowe, Yi~I Wu, Aviv Tamar, Jean Harb, OpenAI Pieter~Abbeel, and Igor Mordatch.
\newblock Multi-agent actor-critic for mixed cooperative-competitive environments.
\newblock \emph{Advances in neural information processing systems}, 30, 2017.

\bibitem[Malik et~al.(2019)Malik, Pananjady, Bhatia, Khamaru, Bartlett, and Wainwright]{malik2019derivative}
Dhruv Malik, Ashwin Pananjady, Kush Bhatia, Koulik Khamaru, Peter Bartlett, and Martin Wainwright.
\newblock Derivative-free methods for policy optimization: Guarantees for linear quadratic systems.
\newblock In \emph{The 22nd International Conference on Artificial Intelligence and Statistics}, pages 2916--2925. PMLR, 2019.

\bibitem[Mondal et~al.(2022)Mondal, Agarwal, Aggarwal, and Ukkusuri]{mondal2022approximation}
Washim~Uddin Mondal, Mridul Agarwal, Vaneet Aggarwal, and Satish~V Ukkusuri.
\newblock On the approximation of cooperative heterogeneous multi-agent reinforcement learning (marl) using mean field control (mfc).
\newblock \emph{The Journal of Machine Learning Research}, 23\penalty0 (1):\penalty0 5614--5659, 2022.

\bibitem[Morimoto and Doya(2005)]{morimoto2005robust}
Jun Morimoto and Kenji Doya.
\newblock Robust reinforcement learning.
\newblock \emph{Neural computation}, 17\penalty0 (2):\penalty0 335--359, 2005.

\bibitem[P{\'e}rolat et~al.(2022)P{\'e}rolat, Perrin, Elie, Lauri{\`e}re, Piliouras, Geist, Tuyls, and Pietquin]{perolat2022scaling}
Julien P{\'e}rolat, Sarah Perrin, Romuald Elie, Mathieu Lauri{\`e}re, Georgios Piliouras, Matthieu Geist, Karl Tuyls, and Olivier Pietquin.
\newblock Scaling mean field games by online mirror descent.
\newblock In \emph{Proceedings of the 21st International Conference on Autonomous Agents and Multiagent Systems}, pages 1028--1037, 2022.

\bibitem[Perrin et~al.(2020)Perrin, P{\'e}rolat, Lauri{\`e}re, Geist, Elie, and Pietquin]{perrin2020fictitious}
Sarah Perrin, Julien P{\'e}rolat, Mathieu Lauri{\`e}re, Matthieu Geist, Romuald Elie, and Olivier Pietquin.
\newblock Fictitious play for mean field games: Continuous time analysis and applications.
\newblock \emph{Advances in Neural Information Processing Systems}, 33:\penalty0 13199--13213, 2020.

\bibitem[Perrin et~al.(2021)Perrin, Lauri{\`e}re, P{\'e}rolat, Geist, {\'E}lie, and Pietquin]{perrin2021mean}
Sarah Perrin, Mathieu Lauri{\`e}re, Julien P{\'e}rolat, Matthieu Geist, Romuald {\'E}lie, and Olivier Pietquin.
\newblock Mean field games flock! {T}he reinforcement learning way.
\newblock In \emph{proc. of IJCAI}, 2021.

\bibitem[Recht(2019)]{recht2019tour}
Benjamin Recht.
\newblock A tour of reinforcement learning: The view from continuous control.
\newblock \emph{Annual Review of Control, Robotics, and Autonomous Systems}, 2:\penalty0 253--279, 2019.

\bibitem[Riley et~al.(2021)Riley, Calinescu, Paterson, Kudenko, and Banks]{riley2021utilising}
Joshua Riley, Radu Calinescu, Colin Paterson, Daniel Kudenko, and Alec Banks.
\newblock Utilising assured multi-agent reinforcement learning within safety-critical scenarios.
\newblock \emph{Procedia Computer Science}, 192:\penalty0 1061--1070, 2021.

\bibitem[Sallab et~al.(2017)Sallab, Abdou, Perot, and Yogamani]{sallab2017deep}
Ahmad~EL Sallab, Mohammed Abdou, Etienne Perot, and Senthil Yogamani.
\newblock Deep reinforcement learning framework for autonomous driving.
\newblock \emph{arXiv preprint arXiv:1704.02532}, 2017.

\bibitem[Sargent and Ljungqvist(2000)]{sargent2000recursive}
Thomas~J Sargent and Lars Ljungqvist.
\newblock Recursive macroeconomic theory.
\newblock \emph{Massachusetss Institute of Technology}, 2000.

\bibitem[Simchowitz et~al.(2020)Simchowitz, Singh, and Hazan]{simchowitz2020improper}
Max Simchowitz, Karan Singh, and Elad Hazan.
\newblock Improper learning for non-stochastic control.
\newblock In \emph{Conference on Learning Theory}, pages 3320--3436. PMLR, 2020.

\bibitem[Sun et~al.(2022)Sun, Kim, and How]{sun2022romax}
Chuangchuang Sun, Dong-Ki Kim, and Jonathan~P How.
\newblock Romax: Certifiably robust deep multiagent reinforcement learning via convex relaxation.
\newblock In \emph{2022 International Conference on Robotics and Automation (ICRA)}, pages 5503--5510. IEEE, 2022.

\bibitem[Tembine(2017)]{tembine2017mean}
Hamidou Tembine.
\newblock Mean-field-type games.
\newblock \emph{AIMS Math}, 2\penalty0 (4):\penalty0 706--735, 2017.

\bibitem[Xie et~al.(2021)Xie, Yang, Wang, and Minca]{xie2021learning}
Qiaomin Xie, Zhuoran Yang, Zhaoran Wang, and Andreea Minca.
\newblock Learning while playing in mean-field games: Convergence and optimality.
\newblock In \emph{International Conference on Machine Learning}, pages 11436--11447. PMLR, 2021.

\bibitem[Yardim et~al.(2023)Yardim, Cayci, Geist, and He]{yardim2023policy}
Batuhan Yardim, Semih Cayci, Matthieu Geist, and Niao He.
\newblock Policy mirror ascent for efficient and independent learning in mean field games.
\newblock In \emph{International Conference on Machine Learning}, pages 39722--39754. PMLR, 2023.

\bibitem[Yongacoglu et~al.(2022)Yongacoglu, Arslan, and Y{\"u}ksel]{yongacoglu2022independent}
Bora Yongacoglu, G{\"u}rdal Arslan, and Serdar Y{\"u}ksel.
\newblock Independent learning and subjectivity in mean-field games.
\newblock In \emph{2022 IEEE 61st Conference on Decision and Control (CDC)}, pages 2845--2850. IEEE, 2022.

\bibitem[Zaman et~al.(2020)Zaman, Zhang, Miehling, and Bașar]{zaman2020reinforcement}
Muhammad Aneeq~uz Zaman, Kaiqing Zhang, Erik Miehling, and Tamer Bașar.
\newblock Reinforcement learning in non-stationary discrete-time linear-quadratic mean-field games.
\newblock In \emph{2020 59th IEEE Conference on Decision and Control (CDC)}, pages 2278--2284. IEEE, 2020.

\bibitem[Zaman et~al.(2021)Zaman, Bhatt, and Ba{\c{s}}ar]{zaman2021adversarial}
Muhammad Aneeq~Uz Zaman, Sujay Bhatt, and Tamer Ba{\c{s}}ar.
\newblock Adversarial linear-quadratic mean-field games over multigraphs.
\newblock In \emph{2021 60th IEEE Conference on Decision and Control (CDC)}, pages 209--214. IEEE, 2021.

\bibitem[Zaman et~al.(2023{\natexlab{a}})Zaman, Koppel, Bhatt, and Ba{\c{s}}ar]{zaman2023oracle}
Muhammad Aneeq~uz Zaman, Alec Koppel, Sujay Bhatt, and Tamer Ba{\c{s}}ar.
\newblock Oracle-free reinforcement learning in mean-field games along a single sample path.
\newblock In \emph{International Conference on Artificial Intelligence and Statistics}, pages 10178--10206. PMLR, 2023{\natexlab{a}}.

\bibitem[Zaman et~al.(2023{\natexlab{b}})Zaman, Miehling, and Ba{\c{s}}ar]{uz2023reinforcement}
Muhammad Aneeq~Uz Zaman, Erik Miehling, and Tamer Ba{\c{s}}ar.
\newblock Reinforcement learning for non-stationary discrete-time linear--quadratic mean-field games in multiple populations.
\newblock \emph{Dynamic Games and Applications}, 13\penalty0 (1):\penalty0 118--164, 2023{\natexlab{b}}.

\bibitem[Zaman et~al.(2024)Zaman, Lauri{\`{e}}re, Koppel, and Ba{\c{s}}ar]{zaman2024robust}
Muhammad Aneeq~uz Zaman, Mathieu Lauri{\`{e}}re, Alec Koppel, and Tamer Ba{\c{s}}ar.
\newblock Robust cooperative multi-agent reinforcement learning: A mean-field type game perspective.
\newblock \emph{arXiv preprint arXiv:2406.13992}, 2024.

\bibitem[Zhang et~al.(2020{\natexlab{a}})Zhang, Chen, Xiao, Li, Liu, Boning, and Hsieh]{zhang2020brobust}
Huan Zhang, Hongge Chen, Chaowei Xiao, Bo~Li, Mingyan Liu, Duane Boning, and Cho-Jui Hsieh.
\newblock Robust deep reinforcement learning against adversarial perturbations on state observations.
\newblock \emph{Advances in Neural Information Processing Systems}, 33:\penalty0 21024--21037, 2020{\natexlab{a}}.

\bibitem[Zhang et~al.(2020{\natexlab{b}})Zhang, Sun, Tao, Genc, Mallya, and Ba{\c{s}}ar]{zhang2020arobust}
Kaiqing Zhang, Tao Sun, Yunzhe Tao, Sahika Genc, Sunil Mallya, and Tamer Ba{\c{s}}ar.
\newblock Robust multi-agent reinforcement learning with model uncertainty.
\newblock \emph{Advances in neural information processing systems}, 33:\penalty0 10571--10583, 2020{\natexlab{b}}.

\bibitem[Zhang et~al.(2021{\natexlab{a}})Zhang, Hu, and Ba{\c{s}}ar]{zhang2021policy}
Kaiqing Zhang, Bin Hu, and Tamer Ba{\c{s}}ar.
\newblock Policy optimization for {H}$\_2$ linear control with {H}$\_\infty$ robustness guarantee: Implicit regularization and global convergence.
\newblock \emph{SIAM Journal on Control and Optimization}, 59\penalty0 (6):\penalty0 4081--4109, 2021{\natexlab{a}}.

\bibitem[Zhang et~al.(2021{\natexlab{b}})Zhang, Yang, and Ba{\c{s}}ar]{zhang2021multi}
Kaiqing Zhang, Zhuoran Yang, and Tamer Ba{\c{s}}ar.
\newblock Multi-agent reinforcement learning: A selective overview of theories and algorithms.
\newblock \emph{Handbook of Reinforcement Learning and Control}, pages 321--384, 2021{\natexlab{b}}.

\bibitem[Zhang et~al.(2021{\natexlab{c}})Zhang, Zhang, Hu, and Ba\c{s}ar]{zhang2021derivative}
Kaiqing Zhang, Xiangyuan Zhang, Bin Hu, and Tamer Ba\c{s}ar.
\newblock Derivative-free policy optimization for linear risk-sensitive and robust control design: Implicit regularization and sample complexity.
\newblock \emph{Advances in Neural Information Processing Systems}, 34:\penalty0 2949--2964, 2021{\natexlab{c}}.

\bibitem[Zhang and Ba{\c{s}}ar(2023)]{zhang2023revisiting}
Xiangyuan Zhang and Tamer Ba{\c{s}}ar.
\newblock Revisiting {LQR} control from the perspective of receding-horizon policy gradient.
\newblock \emph{IEEE Control Systems Letters}, 2023.

\bibitem[Zhang et~al.(2023)Zhang, Hu, and Ba{\c{s}}ar]{zhang2023learning}
Xiangyuan Zhang, Bin Hu, and Tamer Ba{\c{s}}ar.
\newblock Learning the kalman filter with fine-grained sample complexity.
\newblock \emph{arXiv preprint arXiv:2301.12624}, 2023.

\bibitem[Ziegler et~al.(2019)Ziegler, Stiennon, Wu, Brown, Radford, Amodei, Christiano, and Irving]{ziegler2019fine}
Daniel~M Ziegler, Nisan Stiennon, Jeffrey Wu, Tom~B Brown, Alec Radford, Dario Amodei, Paul Christiano, and Geoffrey Irving.
\newblock Fine-tuning language models from human preferences.
\newblock \emph{arXiv preprint arXiv:1909.08593}, 2019.

\end{thebibliography}
